\documentclass[12pt,preprint]{aastex}
\usepackage{graphicx}
\usepackage{epsfig}
\usepackage{multirow}
\usepackage{footnote}

\shorttitle{Decaying Planetary Systems}
\shortauthors{C. Melis et al.}

\begin{document}


\title{Echoes of a decaying planetary system: the gaseous and dusty disks surrounding three white dwarfs}


\author{C. Melis\altaffilmark{1,2}, M. Jura\altaffilmark{1}, L. Albert\altaffilmark{3}, B. Klein\altaffilmark{1}, B. Zuckerman\altaffilmark{1}}
\email{cmelis@ucsd.edu}


\altaffiltext{1}{Department of Physics and Astronomy, University of California,
Los Angeles, CA 90095-1547, USA}
\altaffiltext{2}{Current address: Center for Astrophysics and Space Sciences, University of California, San Diego, CA 92093-0424, USA}
\altaffiltext{3}{CFHT Resident Astronomer, 65-1238 Mamalahoa Highway, Kamuela, HI 96743, USA}


\begin{abstract}
We have performed a comprehensive ground-based observational
program aimed at characterizing the circumstellar material
orbiting three single white dwarf stars previously known to possess 
gaseous disks. Near-infrared imaging unambiguously
detects excess infrared emission towards
Ton 345 and allows us to refine models
for the circumstellar dust around all three white dwarf stars.
We find that each white dwarf hosts gaseous and dusty disks
that are roughly spatially coincident, a result that is consistent
with a scenario in which dusty and gaseous material has its
origin in remnant parent bodies of the white dwarfs' planetary
systems. We briefly describe a new model for the gas disk
heating mechanism in which the gaseous material behaves like a 
``Z~II'' region. In this Z~II region, gas primarily composed of metals is
photoionized by ultraviolet light and cools through 
optically thick allowed Ca~II-line emission.
\end{abstract}


\keywords{circumstellar matter --- planet-star interactions --- stars: individual (Ton 345, SDSS J122859.93+104032.9, SDSSJ104341.53+085558.2) --- white dwarfs}



\section{Introduction}

White dwarfs represent the endstate of stellar evolution for the majority
of stars within our galaxy. Most stars with mass less than $\sim$8 M$_{\odot}$ 
will end up as these exposed cores of nuclear ash that slowly 
cool over time as they radiate away their energy. The planetary
systems that may have orbited these white dwarfs while they were on the
main sequence have been shown to be able to survive through
the asymptotic giant branch (AGB) phase of stellar evolution
\citep{sackmann93,duncan98,debes02,burleigh02}. What is the ultimate
fate of these remnant planetary systems?

The work of \citet{duncan98} showed numerically that giant planets orbiting 
white dwarfs descendent from massive progenitor stars are likely to become 
dynamically unstable on timescales
less than $\sim$10 Gyr. Less massive progenitor stars are more likely to have giant
planets with stable orbits at the end of their stellar evolution. 
However, these less massive progenitor stars are also expected to
obliterate any terrestrial-like planets orbiting within $\sim$1 AU
during their AGB evolutionary
phase, whereas all terrestrial planets
around more massive stars will likely survive through the AGB
\citep[e.g.,][]{sackmann93}.
Rocky objects; like comets, asteroids, or terrestrial planets; 
are expected to survive to the white dwarf
evolutionary phase for a variety of initial sizes and orbital semi-major
axes \citep{jura08}, but may suffer orbital
de-stabilization through the perturbations of giant planets and be flung 
towards the white dwarf star \citep{debes02}. 

Once within the white dwarf's Roche
radius, rocky bodies will be tidally shredded and eventually accrete onto
the white dwarf's photosphere 
\citep{debes02,jura03a,zuckerman07,klein10,dufour10}. 
Sub-micron size grains orbiting within
the Roche radius of their host white dwarf star can produce a
noticeable infrared excess above what would be expected from the white
dwarf photospheric emission alone. Directly detecting
rocky objects in orbit around white dwarfs will be exceedingly difficult;
in contrast, the dusty finale of rocky objects that fall into 
a white dwarf's Roche radius and the eventual atmospheric pollution from
the residual rocky object material leave a tell-tale sign that
there is a dissolving remnant planetary system. Thus, much like 
current searches around main sequence stars
utilize debris disks as signposts for planetary systems 
\citep[e.g.,][]{zuckerman04d,krivov10}, remnant rocky planets around 
white dwarfs are currently best investigated indirectly through searches
for white dwarf debris disks and atmospheric pollution.

Past searches for remnant planetary systems around white dwarfs have
focused on direct and indirect detection of giant planet companions
\citep[e.g.,][and references therein]{hogan09,mullally08} and the
detection of dust contained within a white dwarf's Roche radius
as discussed above \citep[e.g.,][and references therein]{farihi09}. 
To date, searches for giant planets have had little success
\citep[but see][]{mullally08,mullally09}. 
The first evidence for a remnant planetary system around a white dwarf
came from the work of \citet{zuckerman87} in which it was discovered
that the nearby white dwarf star G29$-$38 hosted infrared excess.
This infrared excess was interpreted by \citet{zuckerman87} to probably be the
signature of a brown dwarf companion. However, subsequent observations
of G29$-$38 found evidence for a silicate emission feature at $\sim$10 $\mu$m
\citep{graham90,tokunaga90,reach05} 
indicating that the excess originated from a dusty debris disk.
Further ground-based investigation of white dwarfs decades later
revealed three new debris disks \citep{becklin05,kilic05,kilic06,kilic07}. 
It took the launching of the mid-infrared optimized
Spitzer Space Telescope \citep{werner04} to obtain the necessary sensitivities
to probe for a statistical sample of debris disk-hosting white dwarfs. Spitzer
proved excellent at discovering debris disks around white dwarf stars,
increasing the sample size by nine \citep[see Table 1 in][and references therein]{farihi09}.
Spitzer observations that reached white dwarf photospheric flux levels
enabled a statistical measure of the incidence of white dwarf debris 
disks.
\citet{farihi09} compiled all previous Spitzer observations and found that
$\sim$1-3\% of all single white dwarfs with cooling ages $\lesssim$0.5 Gyr
(T$_{\rm eff}$$\sim$10,000-20,000 K) 
have debris disks and that $\sim$50\% of white dwarfs with implied metal 
accretion rates of $\dot{M}>$3$\times$10$^{8}$ g s$^{-1}$ 
have dusty material
orbiting them. All white dwarfs discussed above have debris disks
contained almost entirely within their respective white dwarf's Roche radius indicating
disrupted rocky objects.

Contemporaneous with the Spitzer white dwarf disk searches, 
\citet{gaensicke06} noted peculiar
double-peaked gas emission lines from a hot (T$_{\rm eff}$=22,000 K)
white dwarf observed as part of the 
Sloan Digital Sky Survey \citep[SDSS;][]{york00}. This single white dwarf,
dubbed SDSS 1228+1040 (spectral type DAZ), was 
interpreted to have an orbiting gaseous metal disk
that had its origin in disrupted asteroidal
material. Searches through the rest of the SDSS white dwarf
spectroscopic database uncovered two more gas-disk white dwarfs 
\citep[SDSS1043+0855, DAZ, and SDSS0843+2257, DBZ;][hereafter
SDSS1043 and Ton 345]{gaensicke07,gaensicke08}. Spitzer and ground-based
observations of SDSS 1228+1040\footnote{\citet{farihi10} performed 
AKARI observations of Ton 345 that indicate there is excess infrared 
emission at 2.3, 3.2, and 4.3 $\mu$m; these observations are contaminated
by a nearby background object discussed in Section \ref{secobs}.}
(hereafter SDSS1228)
show that this gas disk-hosting white dwarf
is also orbited by dusty material that coincides positionally with the
gaseous material \citep{brinkworth09}. Such a discovery heralds the use
of such gas and dust disk hosting white dwarfs as benchmark objects for
studying the decay of planetary systems as the two disk components
yield complementary insights; for example, gas disk measurements 
can directly inform one about the dynamics of the disk while dust disk 
modeling indirectly provides information about disk structure.

Through judicious observing with both ground- and space-based facilities there
is now convincing evidence that white dwarfs show signs of dissolving
planetary systems. However, the sample size of debris disk white dwarfs is
still small and would benefit from additional objects. Furthermore, it is not yet
clear how the gas disk white dwarfs fit into the dissolving planetary system
framework that explains the white dwarf debris disk observations. 
Thus, in an effort to better understand
the final throes of planetary system evolution, ground-based near-infrared and
optical spectroscopic observations of the three gas-disk systems discovered by
\citet{gaensicke06,gaensicke07,gaensicke08} were performed and are reported herein.


\section{Observations}
\label{secobs}

\subsection{WIRCam Imaging at the CFHT}

Initial near-infrared (NIR) images of Ton 345  and SDSS1043
were obtained with WIRCam mounted on the 3.6-m CFHT telescope 
\citep[][]{puget04} on the nights of 
UT 24 February 2008 and UT 19 June 2008, respectively. 
We used a 6- to 9-point dither pattern with 60$\arcsec$ offsets between 
dithers. For Ton 345 we took exposures in the 
J, H, and K$_{\rm s}$ bands with 45, 15, and 
25 seconds per dither (yielding total integration times of 270, 630, and 450 
seconds, respectively). 
The H-band observations for Ton 345 were repeated on UT 25 February 2008 
in better sky brightness conditions.
Both H-band data sets yielded the same result.
For SDSS1043 we took exposures only in the K$_{\rm s}$-band; these
observations were performed in a manner 
similar to that described for Ton 345.
Images were preprocessed, non-linearity corrected, and sky subtracted at CFHT 
with the `I`iwi pipeline
\begin{footnote}
\verb+http://www.cfht.hawaii.edu/Instruments/Imaging/WIRCam/IiwiVersion1Doc.html+
\end{footnote}
and median-stacked using the suite of Terapix software ({\sf sextractor} -
\citet{bertin96}, {\sf scamp} - \citet{bertin06}, and 
{\sf swarp}
\begin{footnote}
\verb+http://www.astromatic.net/+
\end{footnote}) 
with the 2MASS
point-source catalogue as the reference for external astrometry. Photometry of 
the two white dwarfs and all 2MASS sources was performed using the \verb+FLUX_AUTO+ 
measurement in SExtractor, version 2.4.4. Absolute photometry was anchored on 
the 2MASS system using a weighted average of all ($>$200) 2MASS point sources in
the full field of view and 5 iterations of 3$\sigma$ outlier rejection. The 
uncertainty on this calibration is $\sim$0.03 magnitudes in all three bands for 
objects with fluxes similar to Ton 345 and $\sim$0.05 magnitudes in K$_{\rm s}$ for
objects similar in flux to SDSS1043.

\begin{figure}
\begin{minipage}[t!]{53mm}
 \includegraphics[width=53mm]{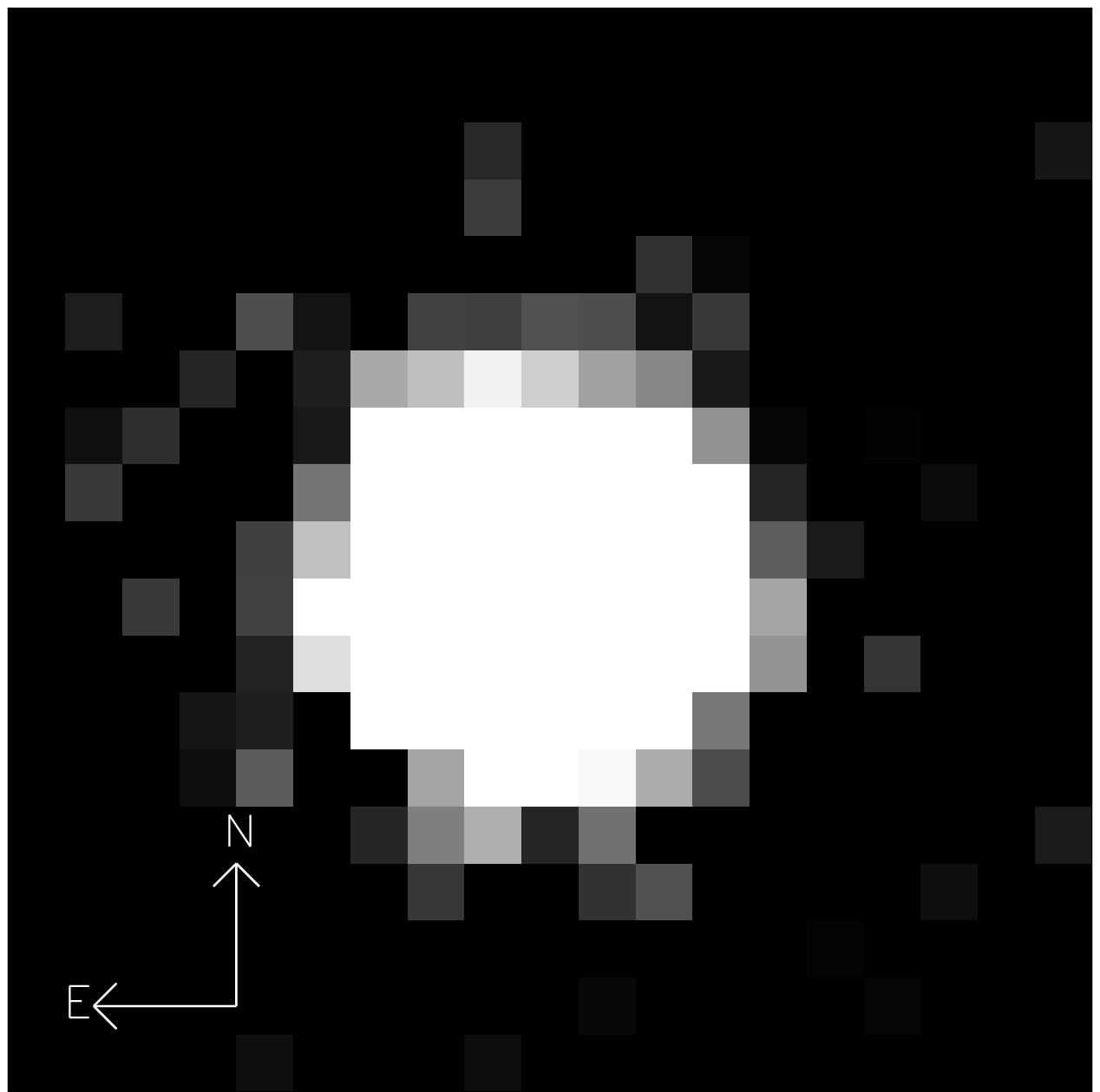}
\end{minipage}
\begin{minipage}[t!]{53mm}
 \includegraphics[width=53mm]{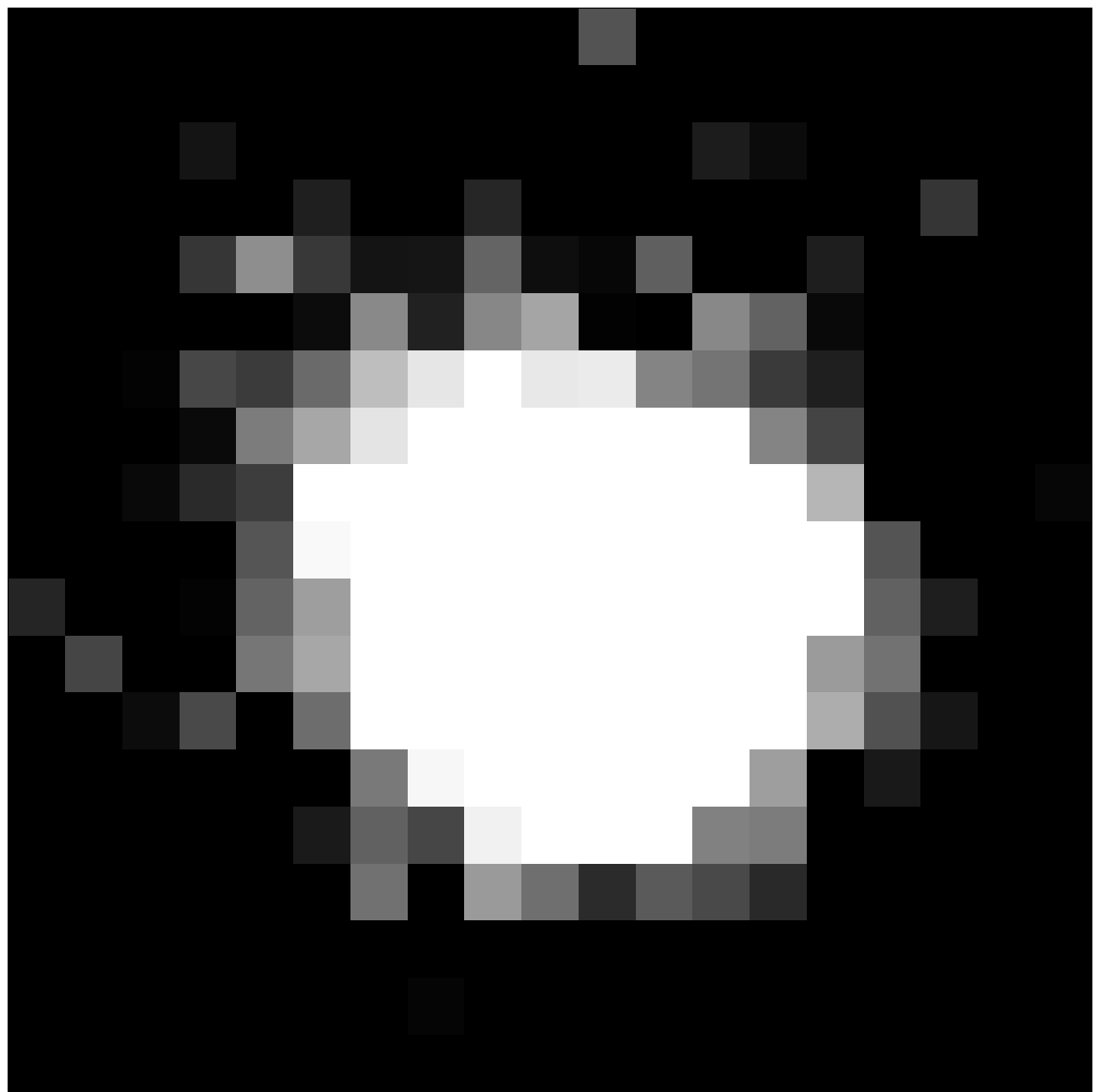}
\end{minipage}
\begin{minipage}[t!]{53mm}
 \includegraphics[width=53mm]{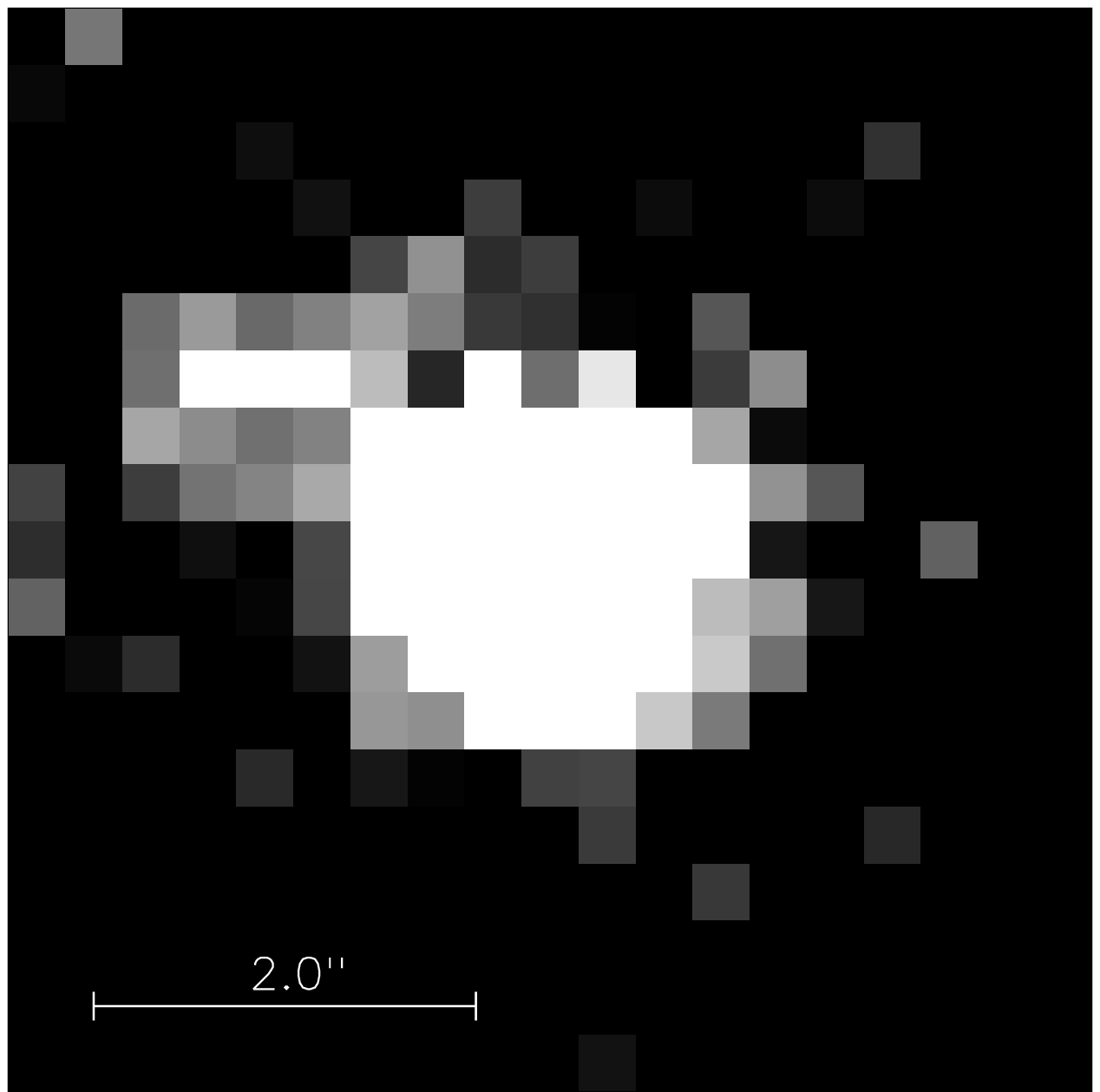}
\end{minipage}
\\*
\begin{minipage}[t!]{53mm}
 \includegraphics[width=53mm]{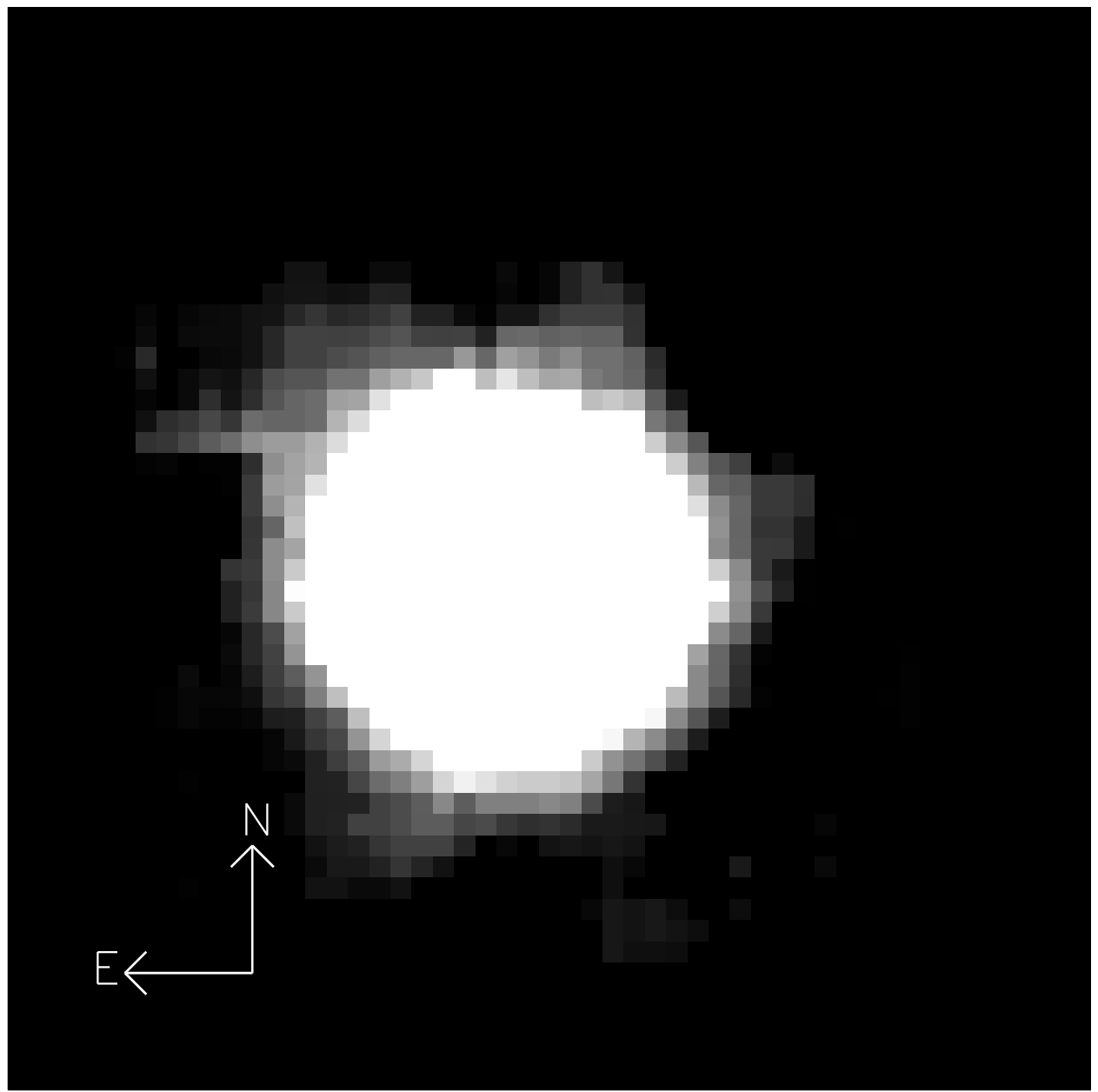}
\end{minipage}
\begin{minipage}[t!]{53mm}
 \includegraphics[width=53mm]{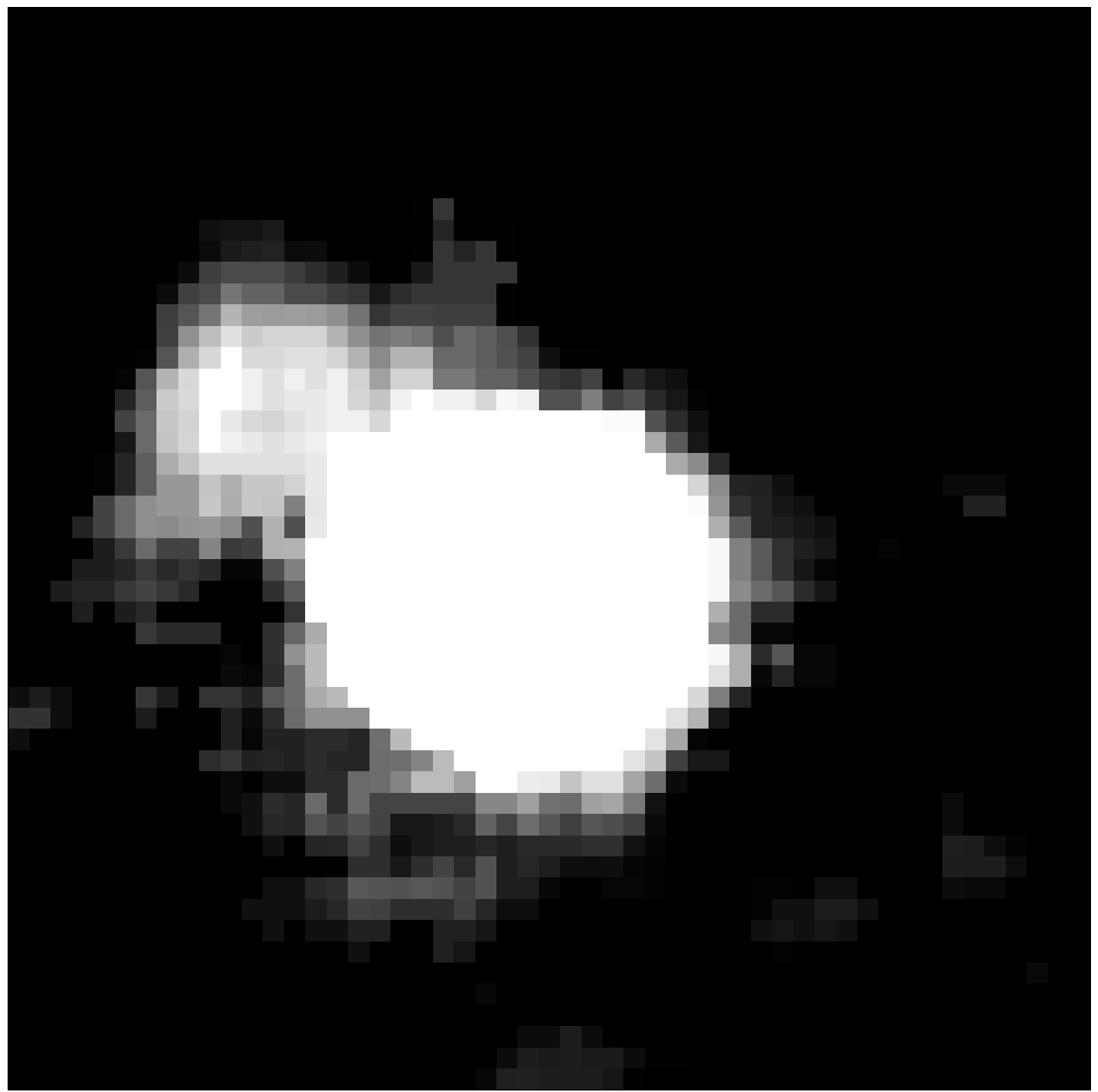}
\end{minipage}
\begin{minipage}[t!]{53mm}
 \includegraphics[width=53mm]{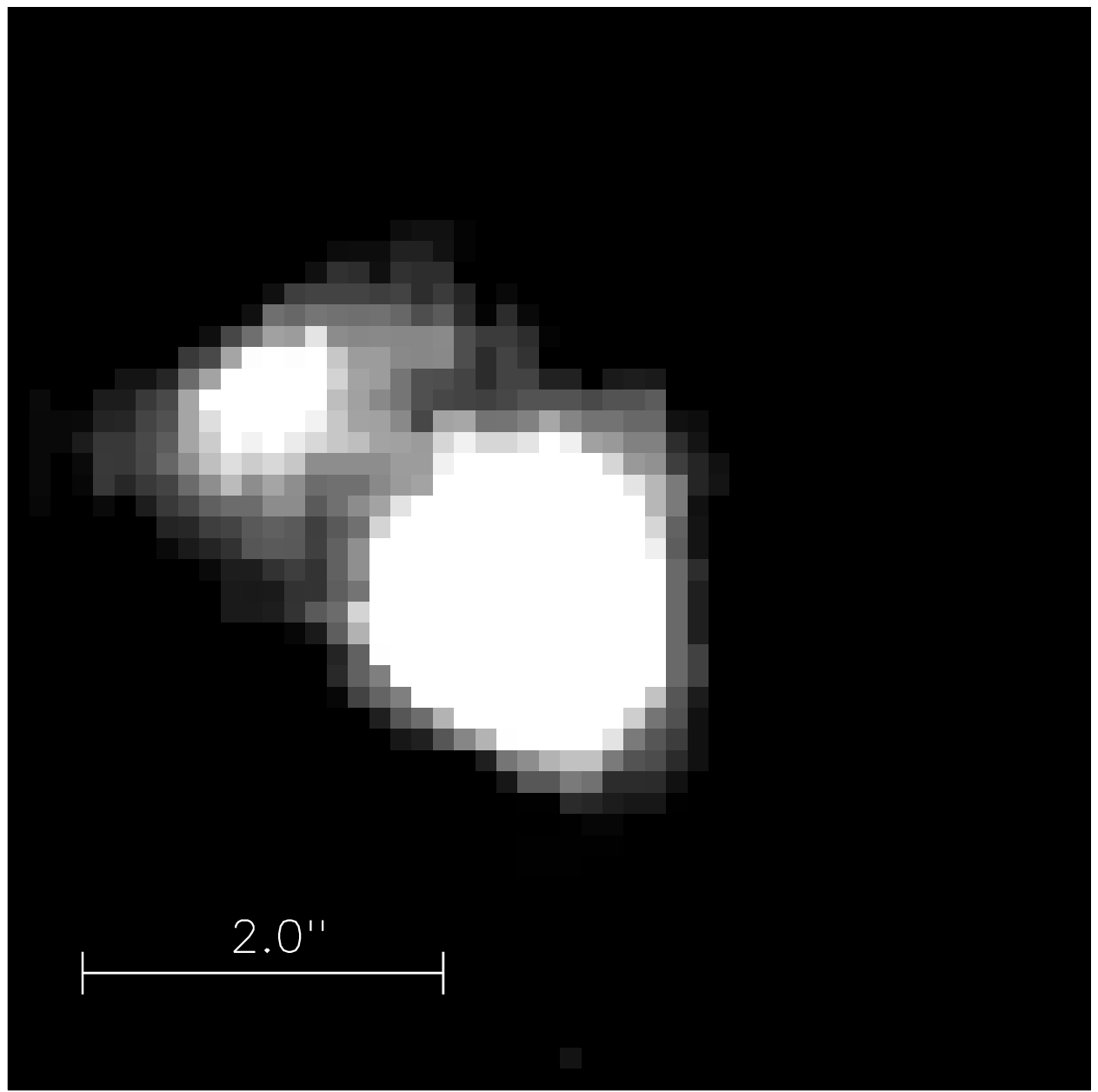}
\end{minipage}
\caption{\label{figjhk} {\it Top panels:} CFHT WIRCam images of Ton 345 in the 
         J-, H-, and 
         K$_{\rm s}$-bands (from left to right) with a linear stretch. The 
         image orientation and 
         scale is the same for all three bands. Note the fainter source located
         $\sim$2.0$\arcsec$ to the NE of Ton 345. This faint background source 
         does not contribute a significant amount of flux to the CFHT measured 
         magnitudes of Ton 345. {\it Bottom panels:} Gemini North NIRI images of
         Ton 345 in the J-, H-, and K$_{\rm s}$-bands (from left to right)
         presented on a linear stretch. Images were smoothed over with a 7-pixel
         boxcar to enhance the faint background object contrast. The image 
         orientation and scale is the
         same for all three bands. The background source, presumably a galaxy,
         is now clearly detected
         in the H- and K$_{\rm s}$-bands.} 
\end{figure}

WIRCam data for Ton 345 are summarized in Table 
\ref{tabtonflux} which also includes SDSS and GALEX fluxes. Images of Ton 345 
and a $\sim$1.8$\arcsec$ separation 
background source, presumably a galaxy, are displayed in Figure \ref{figjhk}. 
Large beam instruments (like Spitzer or AKARI) would be contaminated by this
red background object. The WIRCam
result for SDSS1043 is presented in Table \ref{tab10flux}.




\subsection{NIRI Imaging at Gemini-North}

Follow-up observations of Ton 345 were obtained
with NIRI \citep{hodapp03} mounted on the Gemini North Telescope.
Observations were performed on UT 24 May 2008 during 
Director's Discretionary Time (program GN-2008A-DD-5). 5-point dither 
pattern image sets were repeated for J, H, and K$_{\rm s}$ with total 
integration times of 105, 120, and 150 seconds, respectively. A 4-point dither 
pattern was repeated for L$\arcmin$ until we accrued a total integration time of
2520 seconds.

After pre-processing to correct for the known NIRI non-linearity
\begin{footnote}
\verb+http://staff.gemini.edu/~astephens/niri/nirlin/+
\end{footnote}, data were reduced using in-house IDL software routines. For each filter,
science frames were median
combined to generate a ``sky-median'' frame which was then subtracted from each 
science frame. Sky-subtracted frames were then flat-fielded using exposures
of the illuminated telescope dome for J, H, and K$_{\rm s}$. L$\arcmin$ frames
were flat-fielded using the median stack of all science frames normalized to
unity. Reduced science 
frames were registered by reversing the header-recorded values for the telescope 
offsets between dithers.

Photometric measurements for Ton 345 were calibrated by comparing extracted 
fluxes for each filter to standard 
stars fluxes (Leggett et al 2003, 2006) observed and reduced in a similar 
fashion. For JHK$_{\rm s}$ the standard 
star was FS 15, for L$\arcmin$ it was HD 84800. We note that HD 84800 appears to have a 
$\sim$0.4$\arcsec$ extent in our L$\arcmin$ images; no telescope problems were noted
at the time of observation. 
JHK$_{\rm s}$ fluxes were extracted with an aperture that yielded
$\sim$85\% encircled energy (with a negligible difference between
Ton 345 and FS 15).
For L$\arcmin$ flux calibration we used a 2.0$\arcsec$ diameter aperture to extract 
the flux of HD 84800 and Ton 345. Uncertainties for JHK$_{\rm s}$ fluxes 
were determined by examining the dispersion of the extracted fluxes for each
of the individual reduced science frames. Since this was not possible in our
L$\arcmin$ imaging sets, we instead made ten randomly placed 
extractions on the final,
median-stacked L$\arcmin$ image using the same aperture that was used to extract
the flux for Ton 345. The standard deviation of this set of values was adopted
as the flux measurement uncertainty. The final signal-to-noise ratio (S/N) is 
$\sim$50 for JHK$_{\rm s}$ while for L$\arcmin$ it is $\sim$3-5.

\begin{figure}
 \begin{center}
  \includegraphics[width=80mm]{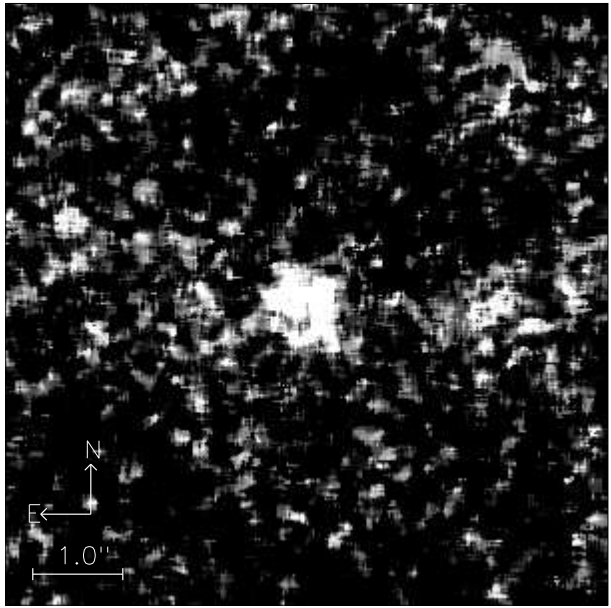}
 \end{center}
\caption{\label{figlp} Gemini North NIRI image of
         Ton 345 in the L$\arcmin$-band presented on a linear stretch after 
         being smoothed over with a 7-pixel boxcar. The fluctuations in the
         background around Ton 345 are representative of the 1$\sigma$ noise 
         level. The background galaxy is not detected, it is likely too diffuse
         for the sensitivity threshold of this imaging set.}
\end{figure}

NIRI photometric results are listed in Table \ref{tabtonflux}. Figures 
\ref{figjhk} and \ref{figlp} show the NIRI images of Ton 345 and the 
$\sim$1.8$\arcsec$ separated galaxy. 
Table \ref{tabgalflux} lists fluxes for the 
galaxy extracted using similar apertures and procedures
as for Ton 345.

\subsection{Gemini Imaging at the Shane 3-m}

Observations of SDSS1043 in the J-band were performed UT 10 October 2008
with the Gemini Twin-Arrays Infrared Camera \citep{mclean93} mounted
on the 3-m Shane telescope at Lick Observatory.
We used a 5 position dither pattern with 20 second integrations
of 5 coadds per position resulting in 500 seconds total on source integration time.
The $\sim$3$^{\prime}$ field-of-view of the Gemini instrument enabled simultaneous
observations of two 2MASS stars for use in flux calibration. Although Gemini
hosts two infrared cameras, data for the longer wavelength chip were unusable
due to instrumental difficulties.

Data were reduced and fluxes extracted much like that described above 
for the Ton 345 NIRI data. Exceptions include no non-linearity
correction, the use of twilight flats instead of dome flats, and image registration
using bright point sources within the field. 
The Gemini J-band flux for SDSS1043 is reported in
Table \ref{tab10flux}.

\subsection{HIRES Spectroscopy}
\label{sechobs}

Keck HIRES \citep{vogt94} optical echelle spectra were obtained for each of the
gas-disk white dwarfs. Table \ref{tabhobs} lists observation dates and HIRES
instrumental setups. Data were reduced via two software packages: IRAF and
MAKEE. Standard IRAF tasks were used in the reduction and extraction of the
HIRES data. After reduction and extraction, low order polynomials were fit to 
each order to bring overlapping order segments into agreement. 

The output of each software package was compared to ensure no reduction
artifacts were contaminating gaseous emission lines. We found that additional 
processing steps were necessary for data taken in the red setup. In some 
observing setups blue light in second order leaked into the $\sim$8200 to 9000 
\AA\ range. This could potentially ``veil'' absorption and emission lines in this
region, especially the Ca~II $\lambda$8662 transition. To estimate the
contaminating flux present we extrapolated the flux from blue contaminant orders 
that fell in between red orders. We believe that this excess flux was removed for
most observations, only the 2008 November observations of SDSS1228 and SDSS1043 
might have significant residuals due to strong Balmer absorption lines that were
difficult to model.

\section{Results}

\subsection{Imaging}
\label{secimg}

Table \ref{tabtonflux} lists measured near-infrared fluxes for Ton 345 from the 
CFHT and Gemini North data sets. To place these near-infrared measurements in context we 
queried SDSS DR7 \citep{abazajian09} and GALEX for {\it ugriz} and ultraviolet fluxes, 
respectively. From the measured magnitudes alone it is apparent that
Ton 345 hosts significant near-infrared excess (see Section \ref{secdust}).

Table \ref{tab10flux} lists GALEX ultraviolet fluxes, SDSS {\it ugriz}, and the
ground-based near-infrared photometry for SDSS1043. Our data alone do not
conclusively show that there is infrared excess emission towards SDSS1043.
But, when these ground-based near-infrared measurements are combined with
Spitzer photometry at longer wavelengths (C.\ Brinkworth {\it et al}.\ 
2010, in preparation),
significant excess infrared emission is apparent (see Section \ref{secdust}).

\subsection{Optical Spectroscopy}
\label{secspec}

\begin{figure}
 \begin{center}
 \begin{minipage}[t!]{83mm}
  \includegraphics[width=83mm]{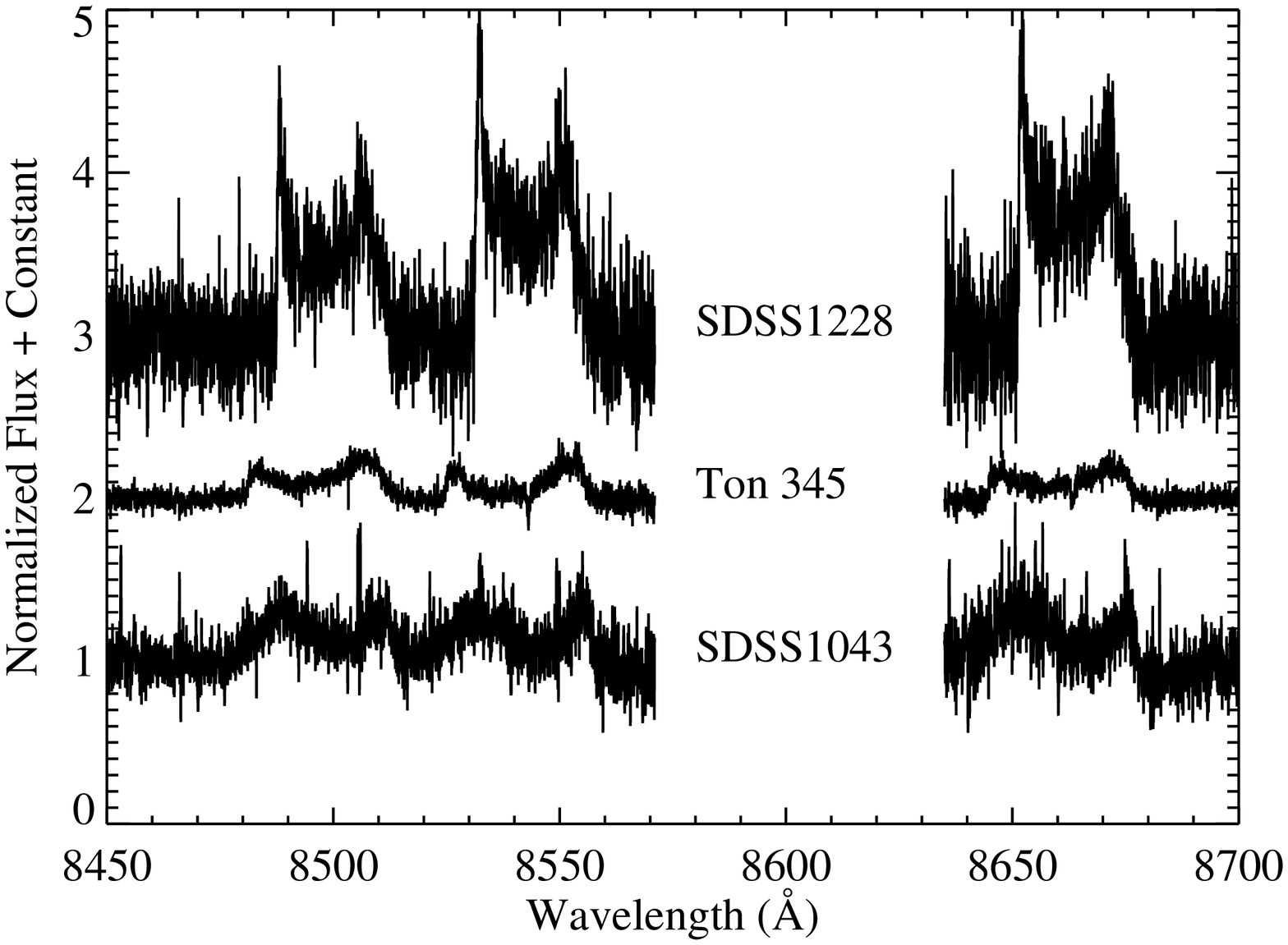}
 \end{minipage}
 \begin{minipage}[h!]{83mm}
  \includegraphics[width=83mm]{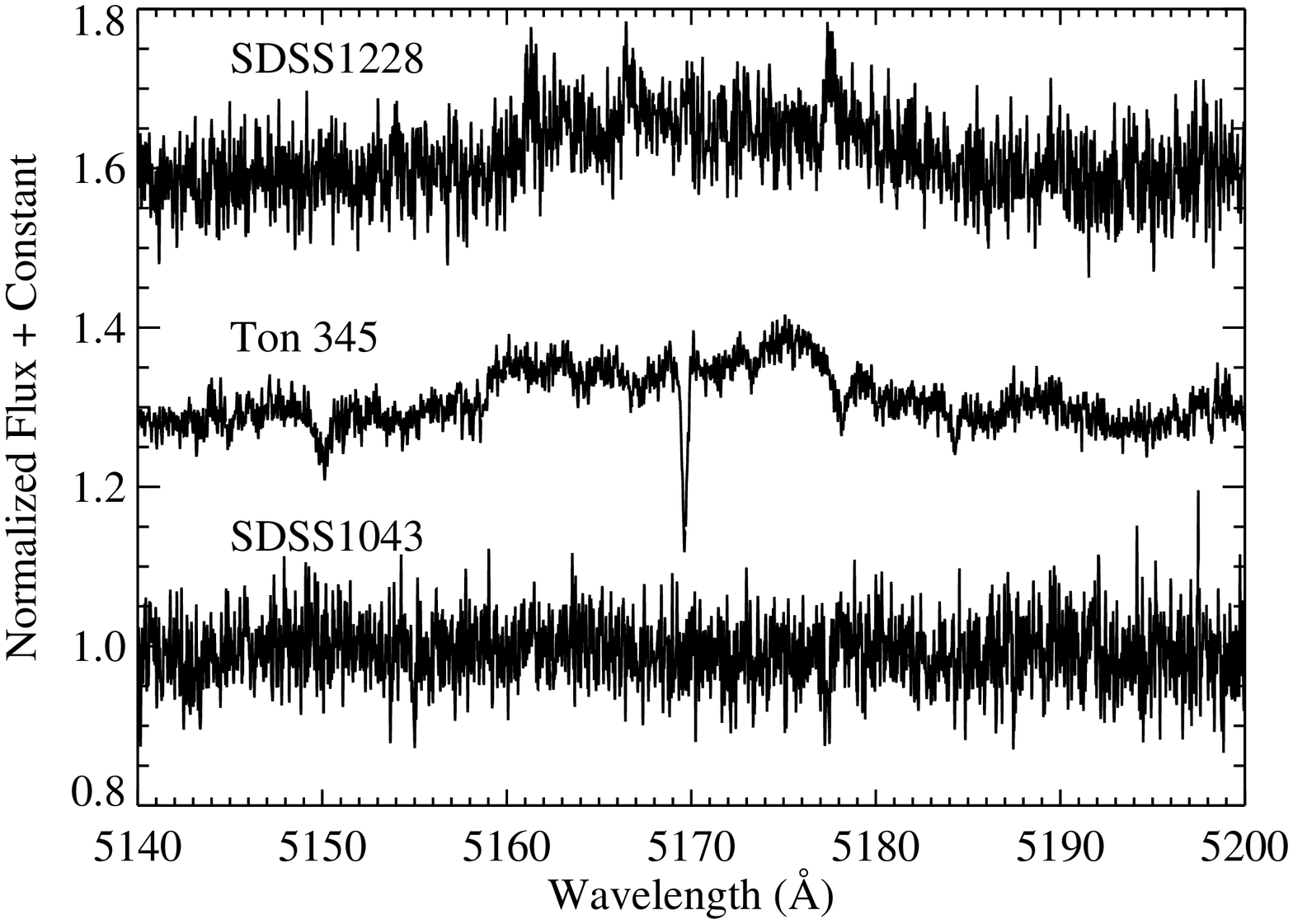}
 \end{minipage}
 \begin{minipage}[b!]{83mm}
  \includegraphics[width=83mm]{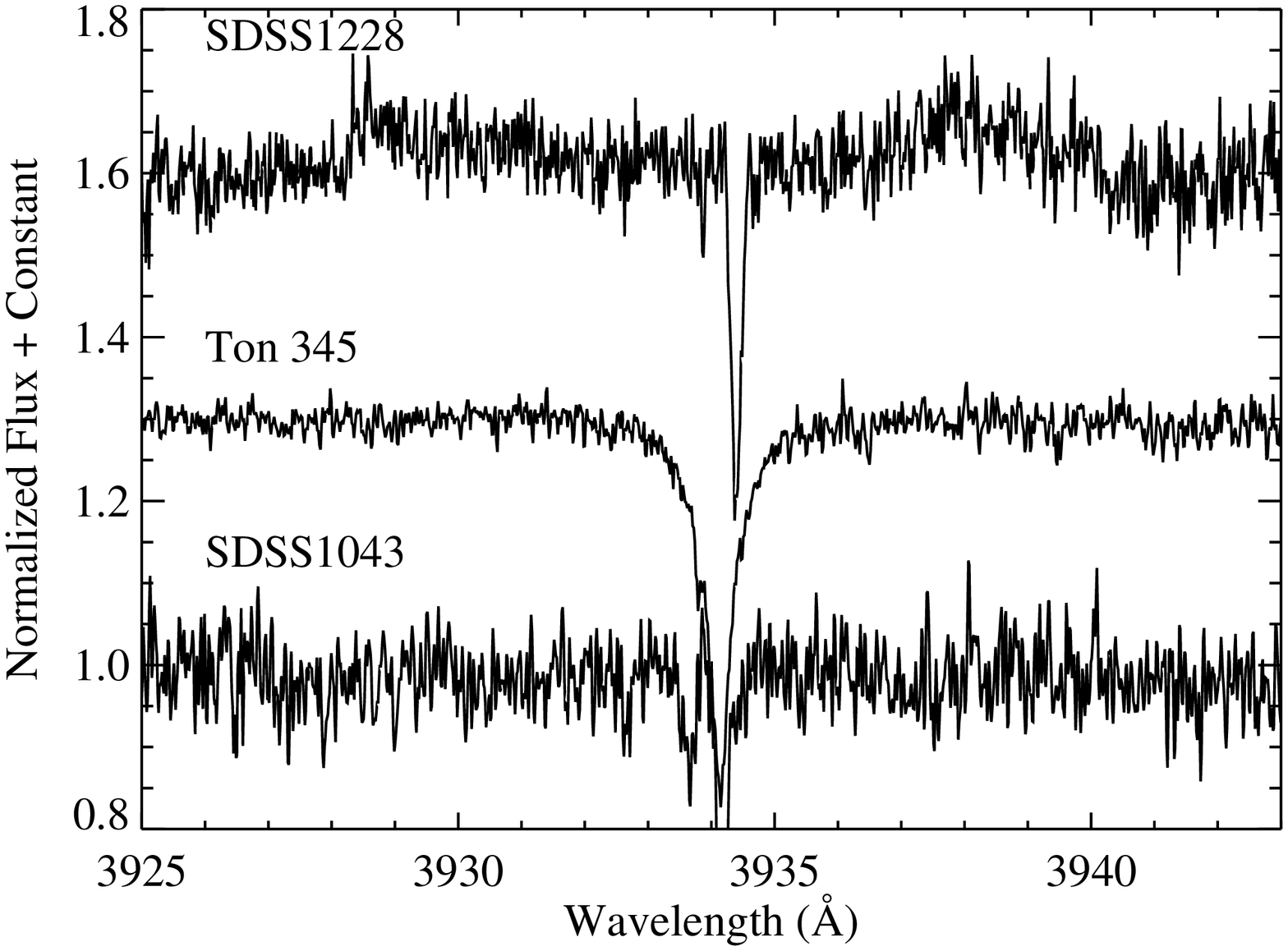}
 \end{minipage}
 \end{center}
 \vskip -0.2in
\caption{\label{fig1gas} \footnotesize{Gas emission lines detected in HIRES optical spectra
         for SDSS1228, Ton 345, and SDSS1043. {\it Top Panel:} Ca~II infrared triplet region. 
         Combined spectra from the 2008 November run are plotted. 
         The lack of data around $\sim$8600 \AA\ is due to the
         gap between red orders for the HIRES setup used. 
         {\it Middle Panel:} Fe~II $\lambda$5169 region. Combined spectra from the 2008
         November run are plotted. Emission lines are seen only in SDSS1228 and Ton 345.
         SDSS1228 appears to have three 
         emission peaks where one may be due to emission from one of the 
         Mg~I~b triplet transitions.
         {\it Bottom Panel:} 
         Ca~II K region. Plotted spectra are from the 2007 May run for SDSS1228 and 
         SDSS1043, and from the 2008 February run for Ton 345. Emission lines 
         are seen only in SDSS1228 (Ca~II H emission for SDSS1228 is detected but not shown).
          The weak, blueshifted Ca~II absorption components
         are likely interstellar in nature based on comparison to known interstellar Ca~II K line
         strengths and velocities in the direction towards each white dwarf \citep{albert93}. 
         Wavelengths in this figure are corrected to the heliocentric
         reference frame and are presented in air.}}
\end{figure}

\begin{figure}
 \begin{center}
 \begin{minipage}[!t]{83mm}
  \includegraphics[width=83mm]{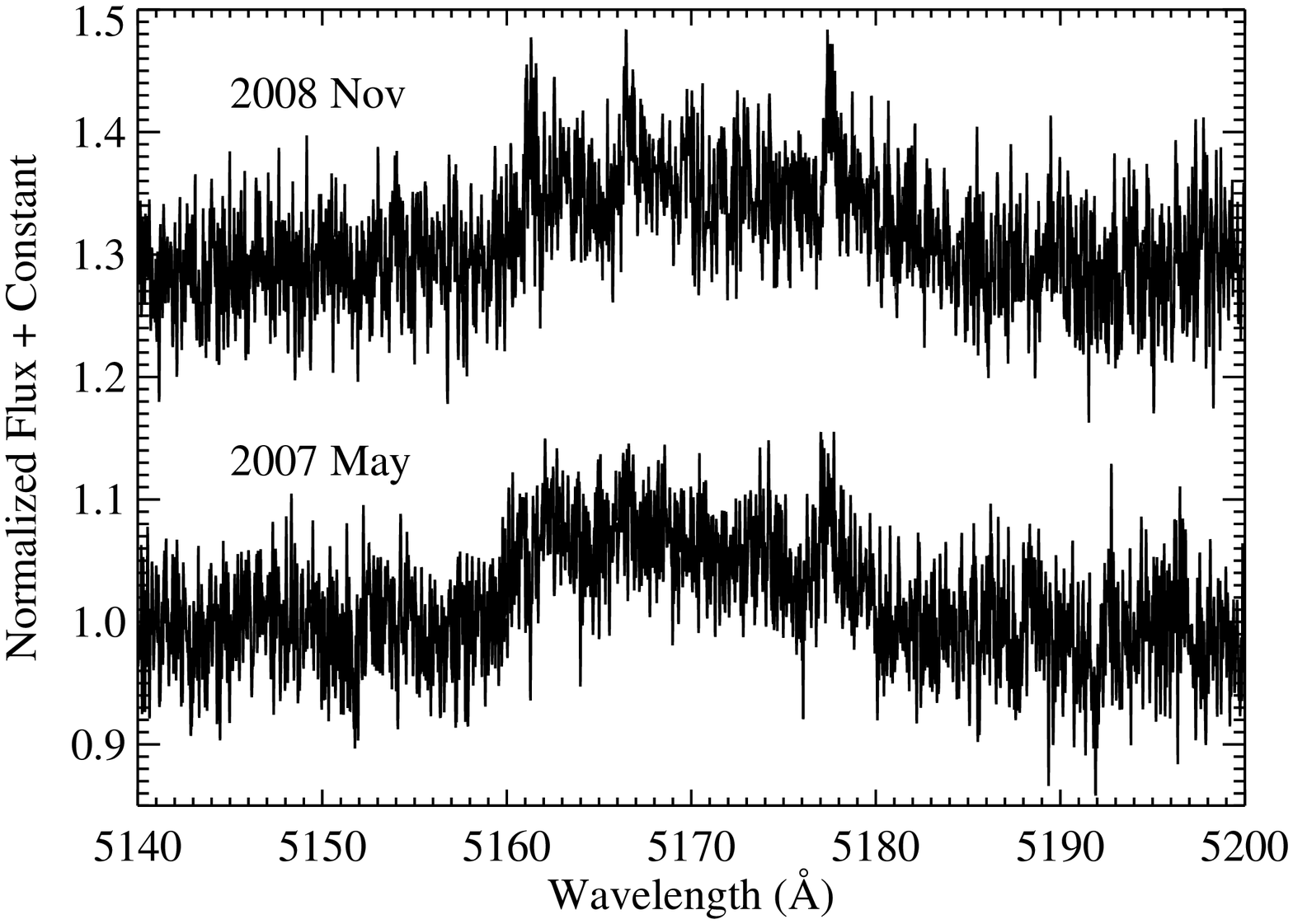}
 \end{minipage}
 \begin{minipage}[!h]{83mm}
  \includegraphics[width=83mm]{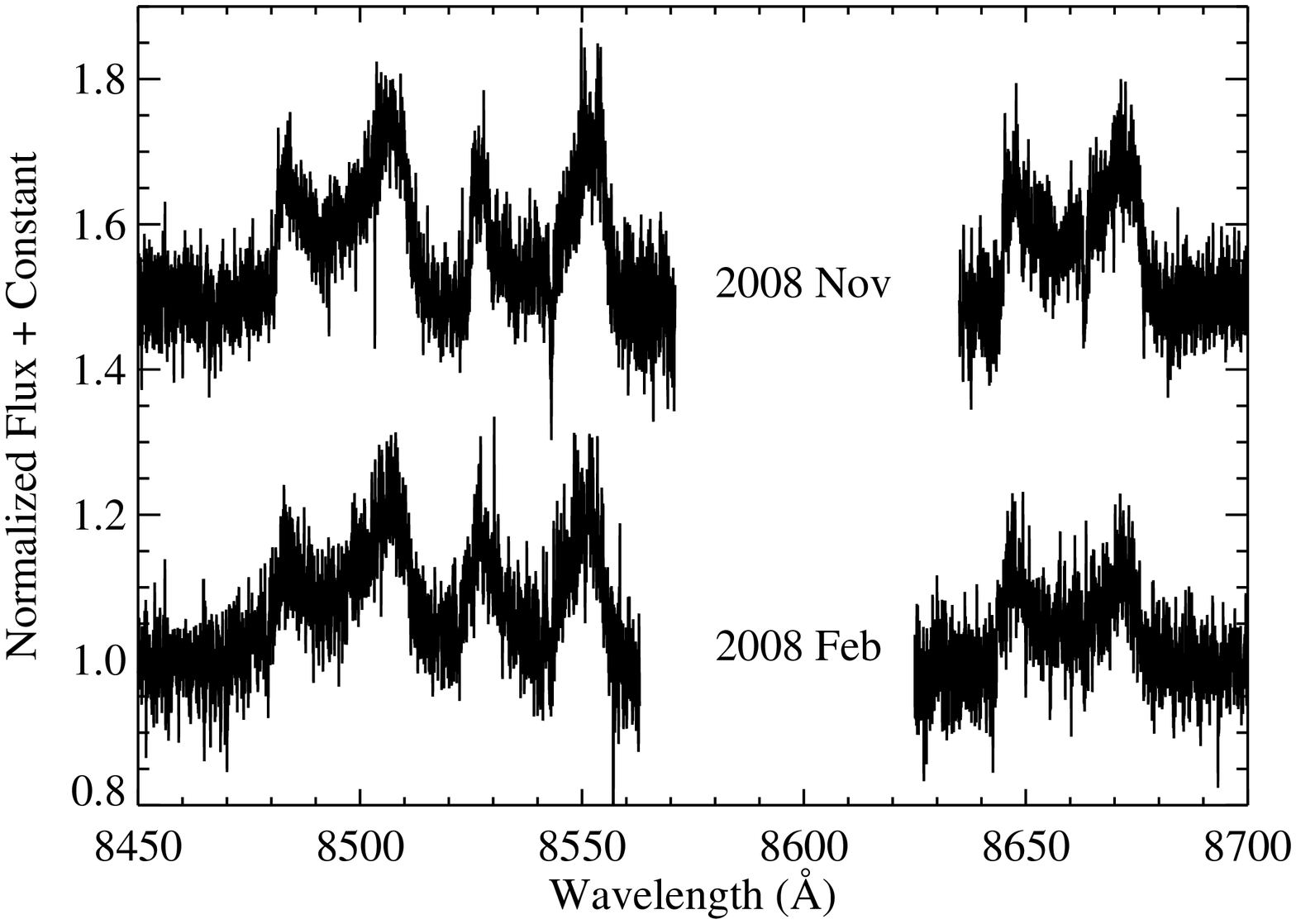}
 \end{minipage}
 \begin{minipage}[!b]{83mm}
  \includegraphics[width=83mm]{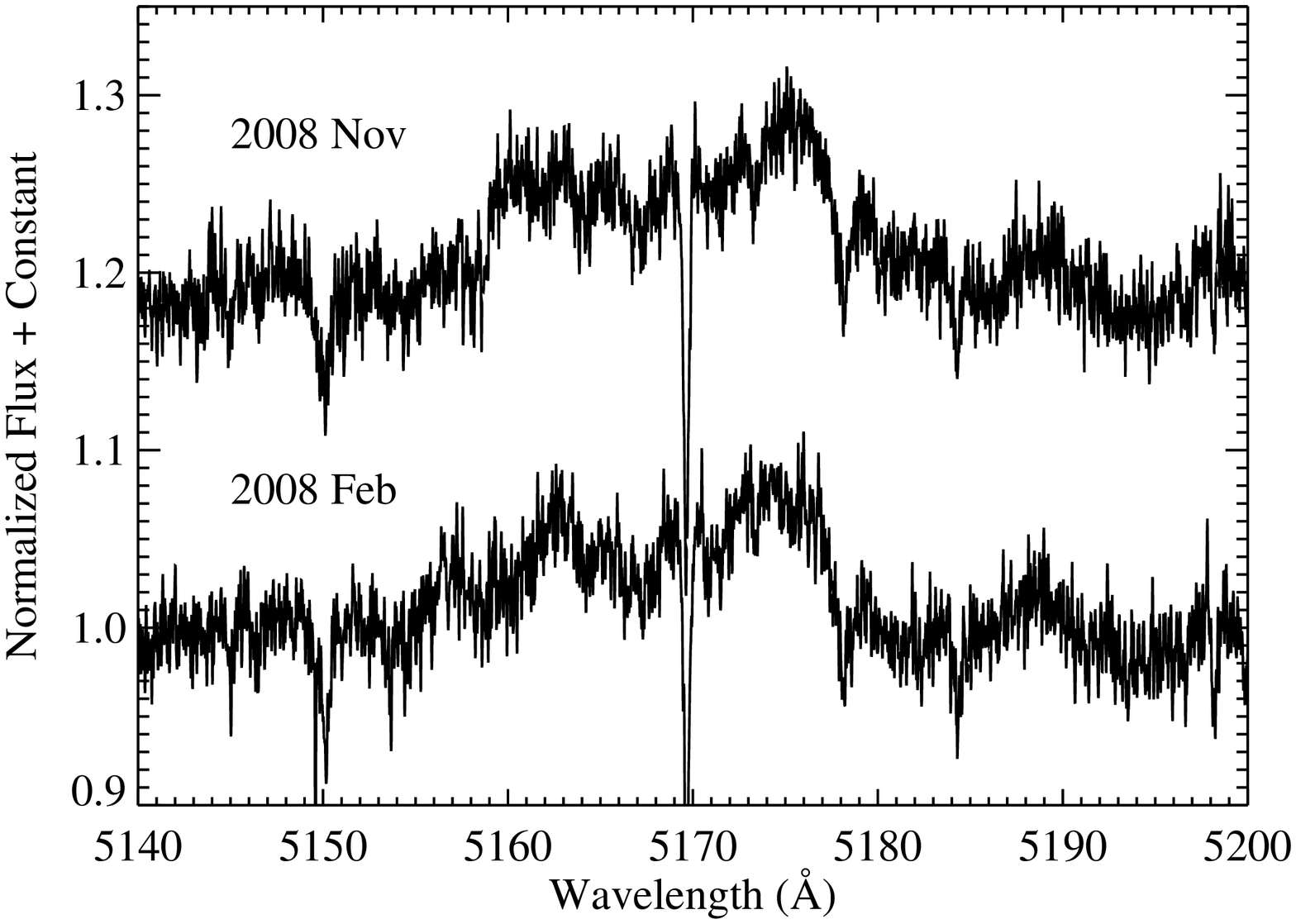}
 \end{minipage}
 \end{center}
 \vskip -0.2in
\caption{\label{fig4gas} \small{Multiple epochs of gas emission lines detected in the HIRES 
         spectra. {\it Top Panel:} Fe~II $\lambda$5169 emission lines for
         SDSS1228. The 2007 May data are from a single exposure. 
         The 2008 Nov data are three combined exposures. The S/N of each 
         are roughly equivalent (see 
         Table \ref{tabhobs}). The relative strength of the three emission peaks seems
         to vary between the two epochs, although the S/N of each epoch
         prevents any conclusive analysis.
         {\it Middle Panel:} Ca~II infrared triplet emission lines for
         Ton 345. All measurable parameters for the two epochs agree to within
          their respective errors.
          {\it Bottom Panel:} Fe~II $\lambda$5169 emission lines for
         Ton 345. All measurable parameters for the two epochs agree to within
          their respective errors. Wavelengths in this figure are corrected to the heliocentric
         reference frame and are presented in air.}}
\end{figure}

Double-peaked metallic gas emission lines are detected towards all three white dwarfs.
Emission lines from Ca~II and Fe~II are shown in Figures
\ref{fig1gas} and \ref{fig4gas}. Ca~II H \& K
emission lines are detected only for SDSS1228 (Figure \ref{fig1gas} and 
Table \ref{tabgas}). Fe~II $\lambda$5169 emission lines are detected 
for SDSS1228 and Ton 345 (Figure \ref{fig1gas}, Figure \ref{fig4gas}, and 
Table \ref{tabgas}). Fe~II $\lambda$5018 emission
lines for SDSS1228 that were noted by \citet{gaensicke06} are not detected. 
Measurable quantities from 
emission features in the spectra, and associated derived quantities, can be found in 
Tables \ref{tabgas} and \ref{tabkin}. We defer quantitative analysis of absorption feature 
equivalent widths and elemental abundances to future publications.

Table \ref{tabgas} contains information on the velocity separation between 
emission peaks of the same transition, 
the full velocity width at zero power of each double-peaked emission feature, 
the ``peak midpoint'', and the maximum velocity gas detected in the blue and
red wings of the double-peaked emission features (v$_{\rm max}$sin$i$). 
We estimate the wavelength centroid position of emission peaks 
by fitting either Gaussian or Lorentzian profiles
over a $\sim$10-15 \AA\ range centered on the emission line peak. Three to 
four fits per peak are performed with the continuum anchor points for the 
fitting function at different locations within the noise of the spectrum. 
The peak centroid is taken to be the average of these multiple fits 
while an estimate of the uncertainty on the centroid comes from the 
standard deviation of these fits. 
The peak midpoint position is then determined by averaging the two emission peak 
centroids for each transition. The peak centroid and peak midpoint
wavelength values are converted into
velocities relative to the rest wavelength for the corresponding transition
they are being measured from.

Measured radial velocities for each star are reported in the
second column of Table \ref{tabkin}. We compute Galactic UVW space 
motions using the gravitational redshift-corrected radial velocities 
\citep[where gravitational redshifts in Table \ref{tabkin} 
are computed using the stellar parameters of][which
are reproduced in
Table \ref{tabkin}]{gaensicke06,gaensicke07,gaensicke08}, 2MASS sky positions
\citep{cutri03},
proper motion information from the NOMAD catalog \citep{zacharias04}, and distances 
computed from the R$_{\rm WD}$/Dist parameter from our model fits 
(see Section \ref{secdust} and Table \ref{tabtonmod}) and the
gravity values given in \citet{gaensicke06,gaensicke07,gaensicke08}. We adopt
an uncertainty of 10\% for the distances.

\subsubsection{Gas Disk Evolution}

With multiple epochs of spectra, and by comparison to
the Ca~II IRT emission lines presented in 
\citet{gaensicke06,gaensicke07,gaensicke08}, we can probe variability in
emission line strength and morphology. Table \ref{tabgas} contains 
emission line equivalent widths, velocity separation between
emission peaks of a given transition, 
and the full width at zero power of emission lines
for each epoch a gas emission line was observed. Figure
\ref{fig4gas} shows multiple epochs of Fe~II $\lambda$5169
emission lines for SDSS1228 and Ton 345 and 
multiple epochs of Ca~II IRT emission in Ton 345. 
There are no significant variations detected in the Fe~II emission lines 
(but see caption to Figure \ref{fig4gas}).

Ton 345's Ca~II IRT emission line equivalent widths and morphologies appear to agree
to within the errors associated with each individual HIRES epoch. This is in 
contrast to the two epochs presented in \citet{gaensicke08} in which the 
equivalent width of the sum of the three Ca~II IRT lines decreases by a factor
of $\sim$1.5 and a significant morphological change in the emission profile is 
seen between 2004 Dec and 2008 Jan. The measurements for the sum of the
Ca~II IRT emission equivalent width reported herein agree with the \citet{gaensicke08} 
2008 Jan value to within the 1$\sigma$ errors. Although speculative, this might
suggest that an episodic disk feeding event occurred before the 2004 Dec
epoch and that the 2008 measurements are probing a ``quiescent'' phase.

Comparing the HIRES observations of SDSS1228's Ca~II IRT complex to those
presented in \citet{gaensicke06,gaensicke07} suggests the stronger of the
two emission peaks has switched from the red side of the double peaked emission
complex \citep[as seen in][]{gaensicke06,gaensicke07} to the blue side (HIRES data).
This parity switch in the peak strength is confirmed
when we degrade the HIRES data to the resolution of the spectra presented in
\citet{gaensicke06}. Interpretation of this parity switch is complicated by the sparse
time sampling of the emission line complexes. The \citet{gaensicke06,gaensicke07}
spectra have epochs of 2003 Mar (SDSS) and 2006 Jul (WHT). 
The morphological appearance of the Ca~II IRT emission lines 
in the SDSS and WHT spectra suggests little change between these
two epochs. It is not clear
what would change the parity in the emission peak strength between the 2006
WHT epoch and the 2008 HIRES epoch. One possibility is that the disk is clumpy
and we have witnessed the clumps orbiting around SDSS1228. Additional 
spectroscopic monitoring of SDSS1228 is necessary to confirm and characterize
such orbiting clumps.

\section{Dust Disk Parameters}
\label{secdust}

\begin{figure}
 \begin{center}
  \includegraphics[width=140mm]{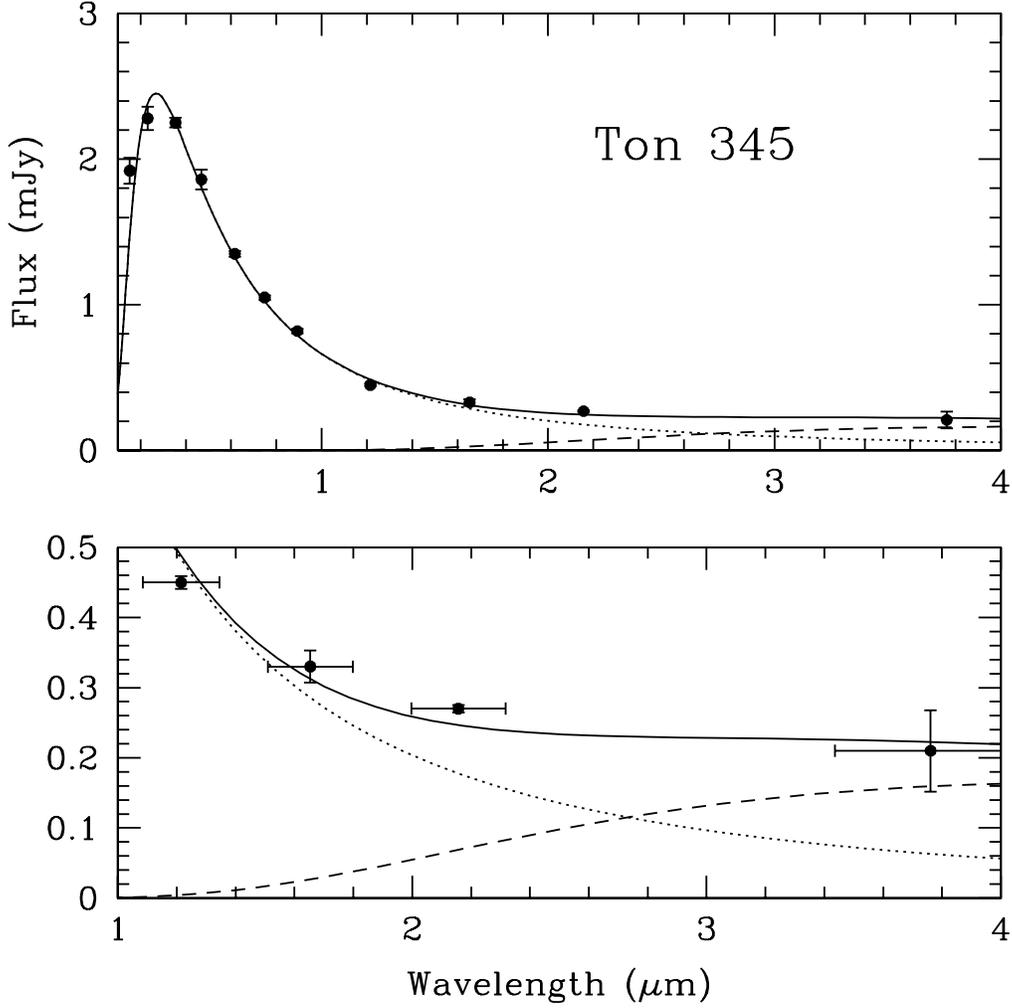}
 \end{center}
\caption{\label{figtonsed} SED for the DBZ white dwarf Ton 345. Overplotted is a model 
         for the white dwarf photospheric  emission 
         (a blackbody with T$_{\rm eff}$$\sim$18,600 K represented by the
         dotted line) and an orbiting 
         flat, passive, opaque dust disk (represented by the dashed line). 
         Broad-band fluxes can be found in Table 
         \ref{tabtonflux} and the model parameters can be found in Table 
         \ref{tabtonmod}. Horizontal bars correspond to filter bandpasses.
         Optical fluxes were obtained from SDSS while ultraviolet 
         fluxes were taken from the GALEX AIS and MIS catalogs. K$_{\rm s}$ and 
         L$\arcmin$ are in excess of what one would expect from the photosphere 
         of the white dwarf alone.}
\end{figure}

Ton 345's photometry suggests near-infrared emission in 
excess of what one would expect from the photosphere of the white dwarf
star alone. The comprehensive data set in Table \ref{tabtonflux} is modeled
as a flat, passive, opaque 
disk \citep{jura03b,jura07b} orbiting a $\sim$18,600 K effective temperature white
dwarf. Ton 345's photospheric emission is assumed to be well represented by a 
blackbody curve; such a model is reasonable for Helium dominated atmosphere white 
dwarfs that do not exhibit strong ultraviolet flux suppression from Hydrogen absorption. 
Figure \ref{figtonsed} displays the white dwarf and disk model
while Table \ref{tabtonmod} reports the model parameters. The photospheric and 
excess fluxes are reproduced succesfully by the white dwarf and disk model. However, 
because the measurements used for Ton 345 only extend out to wavelengths as long
as L$\arcmin$, the temperature at the outer edge of the dust disk is 
observationally unconstrained. In the model shown in Figure \ref{figtonsed} and
Table \ref{tabtonmod} we assume T$_{\rm outer}$ of the disk is 1000 K. 
T$_{\rm outer}$ could be, and likely is, smaller than this value (see below). 
Further observational constraint of the disk outer edge temperature will require 
longer wavelength data that is uncontaminated by the nearby background object.

\begin{figure}
 \begin{center}
  \includegraphics[width=140mm]{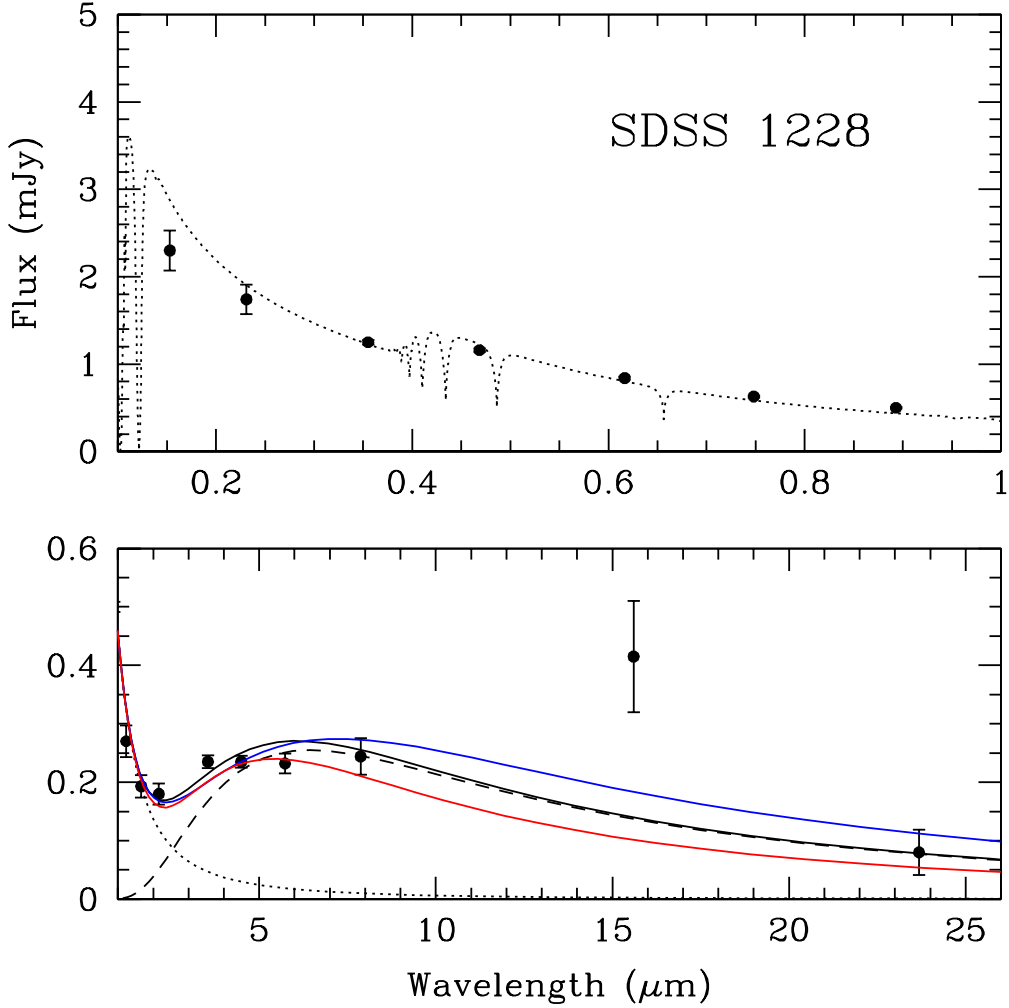}
 \end{center}
\caption{\label{fig12sed} SED for the DAZ white dwarf SDSS1228. Overplotted are models 
         for an atmosphere of a T$_{\rm eff}$=22,020 K, log($g$)=8.24, DA white dwarf
         (dotted line; D. Koester, 2009 private communication) with an orbiting 
         flat, passive, opaque dust disk (dashed line). 
         Input infrared fluxes are taken from \citet{brinkworth09}.
         Three different disk models are displayed to illustrate a
         range of viable model parameters. The black model curve (the middle of the
         three) is the best fit and the model whose parameters we adopt for SDSS1228's dusty
         disk (see Table \ref{tabtonmod}). 
         Optical fluxes were obtained from SDSS while ultraviolet 
         fluxes were taken from the GALEX AIS catalog. One possible explanation for
         the relatively high flux at 16 $\mu$m is that the source possesses a strong silicate
         emission feature \citep{brinkworth09},
         thus it is not included in the continuum fit.}
\end{figure}

\begin{figure}
 \begin{center}
  \includegraphics[width=140mm]{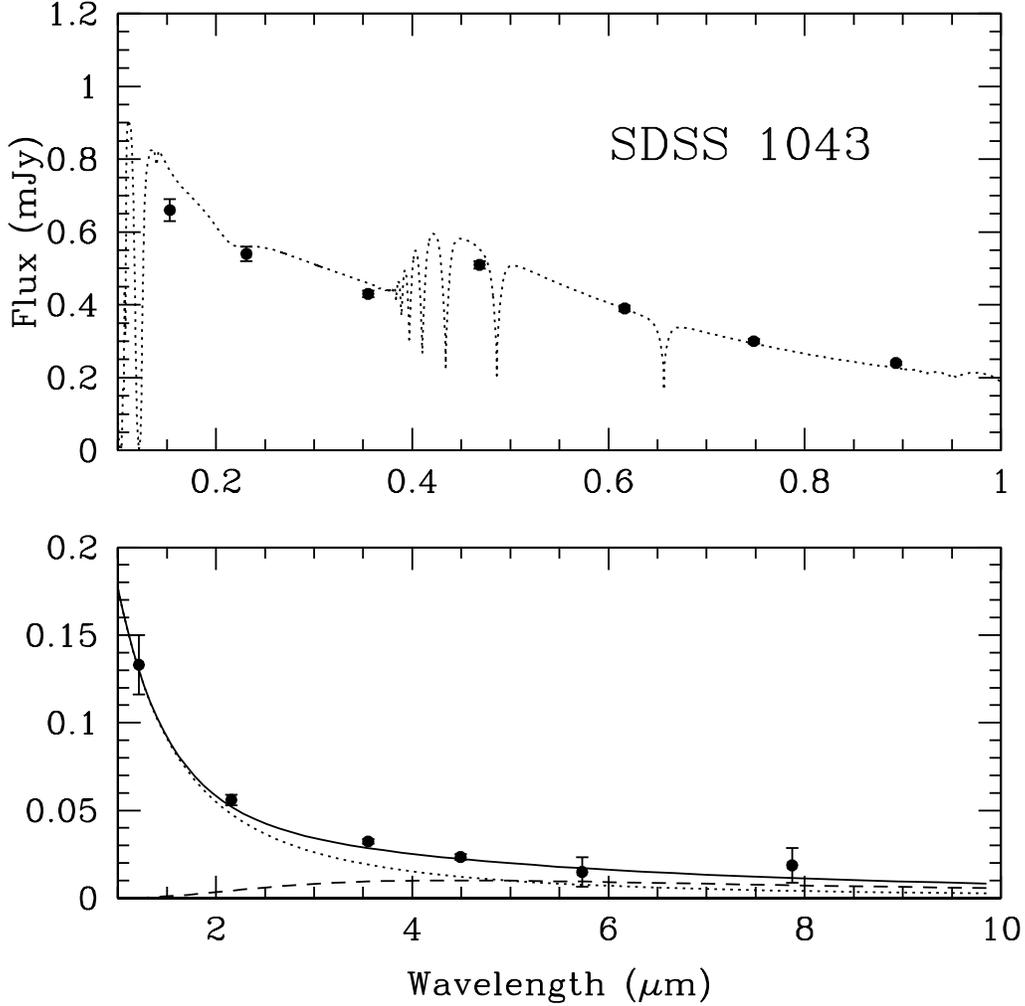}
 \end{center}
\caption{\label{fig10sed} SED for the DAZ white dwarf SDSS1043. Overplotted is a model 
         atmosphere for a T$_{\rm eff}$=18,330 K, log($g$)=8.09, DA white dwarf 
         (dotted line; D. Koester, 2009 private communication) with an orbiting 
         flat, passive, opaque dust disk (dashed line). The model UV fluxes are reddened using the
         \citet{cardelli89} extinction curve assuming E(B$-$V)=0.03. Fluxes can be found in Table 
         \ref{tab10flux} and the model parameters are presented in Table 
         \ref{tabtonmod}. Data points longward of 2 $\mu$m are IRAC measurements
         reported in C.\ Brinkworth {\it et al.} (2010, in preparation). 
         Optical fluxes were obtained from SDSS while ultraviolet 
         fluxes were obtained from the GALEX AIS catalog. When plotted with the 
         Spitzer IRAC data, the K$_{\rm s}$ measurement suggests an excess beginning
         at $\sim$2 $\mu$m.}
\end{figure}

In the interest of comparing a homogeneous set of disk model parameters, we refit
the data for SDSS1228 presented in \citet{brinkworth09} and the data
for SDSS1043 presented in C.\ Brinkworth {\it et al}.\ (2010, in preparation) in the same
manner as Ton 345. These model fits and measured fluxes are shown in 
Figures \ref{fig12sed} and \ref{fig10sed} with model parameters reported
in Table \ref{tabtonmod}. It is noted that SDSS1043's
outer disk temperature is poorly constrained
even with the inclusion of IRAC photometry.
In an effort to illustrate the range of viable model parameters that can 
fit the measured infrared excesses, we plot in Figure \ref{fig12sed}
three models for SDSS1228 that have different dust disk parameters (Table \ref{tabtonmod}).
The resultant parameter extremes reported in Table \ref{tabtonmod} are taken
to be the $\sim$5$\sigma$ confidence
limit on each model parameter.  We then derive the 1$\sigma$ model parameter 
uncertainties by dividing the difference between parameter extremes by five; these
uncertainties are reported
along with SDSS1228's best fit model parameters in Table \ref{tabtonmod}.

For the case of optically thick, flat dust disks Eq.\ 1 from \citet{jura03b} can be 
used to determine at what radial distance from each white dwarf dust 
particles of certain temperatures reside. As previously mentioned,
for Ton 345 and SDSS1043 the temperature of the outer region of 
their orbiting dusty material is poorly constrained by observations. 
Rather than take the model outer disk temperatures for
Ton 345 and SDSS1043 (which are better interpreted as upper limits), 
we instead estimate their
outer disk temperatures under the assumption of an asteroidal
debris model for the origin of the dusty disk. Such a scenario
requires a rocky object to fall within the Roche radius of the white dwarf
\citep{debes02,jura03b}. The following expression for the 
Roche radius is used \citep{davidsson99}:

\vskip -0.3in
\begin{eqnarray*}
 R_{tide} = C_{tide} \left( \frac{\rho_*}{\rho_a}\right)^{1/3} R_* 
\end{eqnarray*}

\noindent where within $R_{tide}$ (the radial separation from a star of radius $R_*$
and density $\rho_*$) a rocky object of density $\rho_a$ would be 
disrupted. The factor $C_{tide}$ is a constant of order unity that is determined
by the tensile strength, shape, rotation rate, and orbital parameters of the
rocky object \citep{davidsson99,holsapple08}. It is unlikely that all objects
that were tidally shredded around SDSS1228, Ton 345, and SDSS1043 had
the same physical parameters; however, in the absence of any such information
it is assumed that this is the case. Such an assumption can potentially introduce
a significant systematic offset in the derived outer dust disk radii.
Using the well-modeld data for SDSS1228
(Table \ref{tabtonmod}) and assuming an asteroid density of $\sim$3 g cm$^{-3}$
a value of 1.45 for $C_{tide}$ is estimated. Combining these values with the
mass and radius of the white dwarfs (see Table \ref{tabkin}) from
\citet{gaensicke06,gaensicke07,gaensicke08} we calculate the tidal radii for 
each white dwarf. These values for Ton 345 and SDSS1043 are reported as 
R$_{outer,dust}$ in Table \ref{tabgasdi}. The value reported for SDSS1228 
in Table \ref{tabgasdi} is the disk outer radius as derived from fitting its Spitzer data.

\section{Gas Disk Parameters}
\label{secgasdiskdi}

\begin{figure}
 \begin{center}
 \includegraphics[width=160mm]{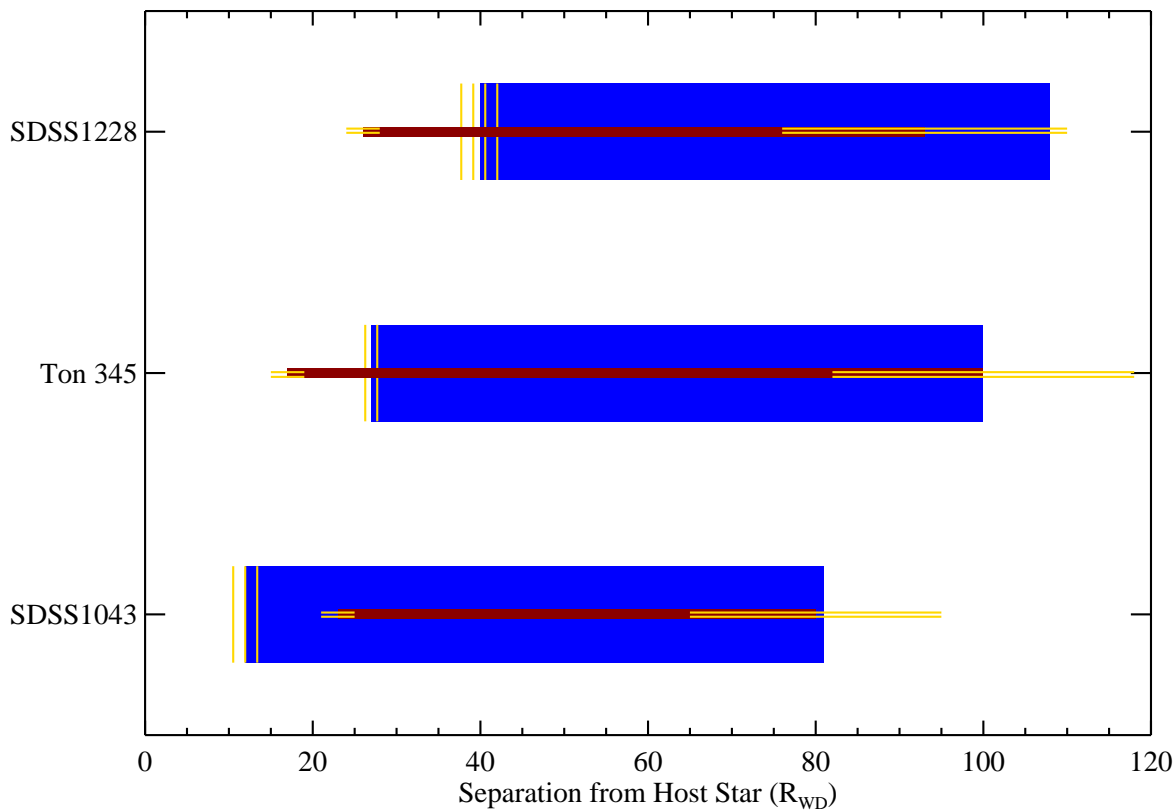}
 \end{center}
\caption{\label{fig7gas} Sketch illustrating the results presented in 
               Table \ref{tabgasdi}. The dimensions of each white dwarf's
               dusty and gaseous disks are plotted in units of each white dwarf's radius. 
               True blue colored regions correspond to gas disks while cardinal red 
               colored regions correspond to dust disks. The  
               vertical scale heights of the gas disks are 
               expected to be larger  
               than those of the dust disks (Section \ref{sec611}). Gold, horizontally hatched regions 
               correspond to the 1$\sigma$ uncertainty for the dust disk inner and outer radii as 
               reported in Table \ref{tabgasdi}. Gold, vertically hatched regions correspond 
               to the 1$\sigma$ uncertainty for the gas disk inner radii as reported in 
               Table \ref{tabgasdi}.}
\end{figure}

Using the HIRES optical spectra we can estimate the radial extent of the
gaseous disks orbiting the three white dwarfs. This is done by measuring the
highest velocity gas emission in the HIRES spectra
(v$_{\rm max}$sin$i$ in Table \ref{tabgas}), emission that corresponds
to the innermost orbit of the emitting gas-phase metals around the white dwarf
\citep[e.g.,][]{horne86}. However,
this value is degenerate with the inclination angle of the disk, $i$, where the
degeneracy is $v$sin$i$ ($i$ of 0$^{\circ}$ would correspond to a face-on disk).
The dust disk model fits (Section \ref{secdust})
can help constrain the disk inclination angle assuming
the dust and gas disks are coplanar. It is noted that the inclination angles as
derived from the dust disk fitting can vary by $\sim$20-30\% from the best-fit value
(5$\sigma$; see Table \ref{tabtonmod} and Section \ref{secdust}).
Line-of-sight gas disk velocities for the three white dwarfs are corrected
using the dust disk model best-fit inclination angles. Assuming Keplerian orbits of the gas
and dust disks, we use the white dwarf masses and radii (see Table \ref{tabkin}) as modeled by
\citet{gaensicke06,gaensicke07,gaensicke08} to derive the gas disk inner
radius in units of white dwarf radii.
We obtain the full dimensions of the gaseous disks by
combining these measurements with the \citet{gaensicke06,gaensicke07,gaensicke08,
gaensicke08ASPC} models (see Section \ref{secmodels})
in which the gas disk outer radii have been estimated. These results
are reported in Table \ref{tabgasdi} and illustrated in Figure \ref{fig7gas}. 

We note the discrepancy between our inferred gas disk
inner radius for SDSS1228 and the same as quoted in \citet{brinkworth09}. The origin of
this discrepancy has to do with two factors.
The first factor is responsible for the smaller gas disk inner radius 
quoted
in \citet{brinkworth09} $-$ R$_{inner,gas}$ $\sim$27 R$_{\rm WD}$ $-$  and
results from use of the maximum Ca~II infrared 
triplet (IRT) gas emission velocity of $\sim$1270 km s$^{-1}$ as
reported in Table 1 in \citet{gaensicke06}. Re-measurement of the spectra
presented in \citet{gaensicke06} provides a maximum gas emission velocity more
in line with that measured herein 
\citep[Table \ref{tabgas}; use of the][erroneous v$_{\rm max}$sin$i$ will reduce the gas disk inner radius by a factor of $\sim$1.8 relative to our inferred gas disk inner radius]{gaensicke06}
and what is quoted in the
supplementary material of \citet{gaensicke06}\footnote{The values reported in the main
article text and tables of \citet{gaensicke06} are different from those reported in the
supplementary online material.}. 
The second factor is
a 20\% increase in the inner disk radius relative to that presented herein
\citep[where the ratio of 1.8/1.2=1.5 is
the ratio between our inferred gas disk inner radius and that of][]{brinkworth09}
and comes from a matter of interpreting Figure 1 of \citet{horne86}. 
The inner gas disk radius reported in Table \ref{tabgasdi}
for SDSS1228 agrees with the value quoted in the supplementary material
\addtocounter{footnote}{-1}
to \citet{gaensicke06}\footnotemark.

\subsection{Stable Eccentric Disk at Ton 345}
\label{secgrav}

Here a novel method of estimating the gravitational redshift for the gas disk-hosting
white dwarf stars is attempted. This method employs the gas disk emission lines
(from which we seek to measure the white dwarf systemic velocity) and 
photospheric absorption lines (which contain velocity components from the
systemic velocity and the gravitational redshift from the compact white dwarf).
To estimate the gravitational redshift we measured the photospheric
absorption line radial velocity for each white dwarf (see RV$_{\rm obs}$ in 
Table \ref{tabkin}) and the systemic radial velocity for each white dwarf-disk system 
from gaseous emission lines (see ``Peak Midpoint Velocity'' in Table \ref{tabgas} 
and discussion in Section \ref{secspec}). Our estimated gravitational redshift for
SDSS1228 (RV$_{\rm obs}$$-$[Peak Midpoint Velocity]$_{\rm avg}$$=$ 55$\pm$5 km s$^{-1}$;
based only on the Ca~II IRT emission lines) 
reproduces well the expected gravitational redshift (44 km s$^{-1}$, see
Table \ref{tabkin}). Estimates for SDSS1043 are inconclusive due to the low S/N in the
gas emission lines. However, our estimate for Ton 345 
(RV$_{\rm obs}$$-$[Peak Midpoint Velocity]$_{\rm avg}$$=$ 131$\pm$6 km s$^{-1}$;
based only on the Ca~II IRT emission lines) is 
quite different from the expected gravitational redshift (44 km s$^{-1}$).

Why does this method fail for Ton 345? A clue comes from considering
the agreement of SDSS1228's gravitational redshifts as measured from the two different
methods. As modeled by \citet{gaensicke06,gaensicke08},
the gas disk orbiting SDSS1228 has negligible eccentricity while for Ton 345 the
disk eccentricity is large ($e$$\sim$0.2$-$0.4). In the case of a large
disk eccentricity it is understandable
that measuring the peak midpoint velocity as described in Section \ref{secspec} would
provide an inaccurate estimate of the systemic velocity. As
such, we instead take the discrepancy between the gravitational redshifts as computed
through the two different methods for Ton 345 as additional evidence for the large
eccentricity of the gaseous disk \citep[$e$$\sim$0.2-0.4, ][]{gaensicke08}. It is noted
that these high eccentricity values in Ton 345's disk persist over long time-spans. 
Ton 345's disk is likely pumped up to (and maintained at) 
high eccentricity by a surviving planet \citep{lubow91,kley06}.

\section{Gas Disk Models}
\label{secmodels}

When observed at high spectral resolution the relatively clean-cut morphologies
of the gaseous emission lines presented in 
\citet{gaensicke06,gaensicke07,gaensicke08} resolve into complex structures. A
common feature of the HIRES observed Ca~II IRT emission lines for all three white dwarfs is a 
profile having the typical peak and wing structure \citep[e.g., see Figure 1 in][]{horne86}
on one side of the emission 
(red side of the double peaked Ca~II emission features for SDSS1228 and Ton 345
and the blue side for SDSS1043 in Figure \ref{fig1gas}) 
while having a peak with a sharp cut-off 
in emission (i.e., no wing) on the other side  (blue side of the double 
peaked Ca~II emission features for SDSS1228 and Ton 345
and the red side for SDSS1043 in Figure \ref{fig1gas})
of the emission. One could imagine that such an emission-line morphology would 
still be well reproduced by modeling the disks with families of elliptical rings as was 
done by \citet{gaensicke06,gaensicke07,gaensicke08}.

\citet{werner09} present a physical model to explain the gas emission lines
observed in SDSS1228's spectrum; their model relies on an active disk that is 
heated by viscous dissipation of disk energy. We use the HIRES
spectra to test the predictions of this model. In particular, we compare the HIRES
spectra covering the Ca~II H, K, and IRT transitions (Figure \ref{fig1gas}) 
to the predictions in Figure 6
of \citet{werner09}. \citet{werner09} predict that the summed Ca~II H \& K emission
complex should have flux that is a factor of $\sim$5 stronger than
the summed Ca~II IRT flux. The Ca~II H, K, and IRT 
emission lines detected in the spectra of SDSS1228 
can be used as a robust test of the \citet{werner09} model. 
For each emission line the continuum flux is subtracted
and the line flux summed. Extraction of the Ca~II H emission 
line flux is not attempted as this complex 
is contaminated by the H$\epsilon$ absorption line. 
We instead assume that the emission line flux is the same for both of the 
Ca~II H \& K transitions 
and multiply the measured Ca~II K emission line flux by 2 as a
proxy for the sum of the emission line flux for both transitions. 
Examination of SDSS1228's emission line flux for the Ca~II H, K, and IRT transitions
indicates that the summed Ca~II H \& K emission line flux
is $\sim$3.5 times weaker than the summed Ca~II IRT emission line flux (see
Table \ref{tablineflux}).  Thus, the observational result differs by a 
factor of $\sim$17 from that expected in the \citet{werner09} model. It is 
noted that the total flux in line emission
is $\sim$3$\times$10$^{-4}$ times the bolometric flux emitted by SDSS1228.

Assuming the line emission is optically thick
\citep[as is determined in the modeling presented in the supplemental material to][]{gaensicke06}, 
we can estimate the gas 
disk temperature from our observations by comparing the emission line 
flux for Ca~II transitions in different parts of the optical spectrum. For SDSS1228,
the only white dwarf with detected Ca~II H+K emission, we 
estimate a gas disk 
temperature of $\sim$5000 K. Gas disk temperature upper limits for Ton 345
and SDSS1043 are reported in Table \ref{tablineflux}.

\subsection{Z~II Region Model}
\label{seczii}

The \citet{werner09} model does not provide a good match to the HIRES data.
As such, we briefly describe an alternative model for the disk
heating mechanism. We envision a model where the gas is photoionized by  
ultraviolet photons from the white dwarf. The idea would be akin to an H~II region, but
since this is a gas that is extremely deficient in H or He
\citep[][and the results preseneted herein]{gaensicke06,gaensicke07,gaensicke08,gaensicke08ASPC}, it would be more aptly
described as a Z~II region. To determine the steady state temperature of a Z~II region
one must consider three processes: 

\begin{itemize}
\item[(i)] injection of energy into the gas from photoionized electrons; 

\item[(ii)] energy loss from the gas when electrons recombine
with ions, a process for Z~II regions that would be negligible when compared
to the last process:

\item[(iii)] energy loss from the gas due to optically thick line emission. 
We assume that the gas density is sufficiently high ($>>$10$^6$ cm$^{-3}$)
that forbidden line cooling is unimportant and that 
gas instead cools through emission lines like the Ca~II IRT.

\end{itemize}

\noindent We expect that the gas disk scale height will be significantly larger
than the dust disk scale height (see Section \ref{sec611}). 
As a result, collisional energy transfer from 
gaseous atomic species to dust particles \citep[e.g.,][]{goldreich74}
will be negligible compared to the above three energy balance points. 
Hence, it is expected that the gas and dust temperatures are decoupled.

Here we provide details for a highly idealized Z~II region model.
We consider a flat passive disk \citep[see][]{jura03b} whose thickness is  
a small fraction of the radius ($R_*$) of the illuminating star. The
star is assumed to have effective temperature 
$T_{*}$ and photospheric emission well described by a blackbody $B_{\nu}(T_{*})$. 
At radial distance 
$D$ from the star,  we compute the effective disk-heating flux of stellar photons 
above energy threshold $h{\nu}_{1}$ with frequencies between ${\nu}_{1}$ and 
${\nu}_{2}$, $F^{-}({\nu}_{1},{\nu}_{2})$ in the limit that $D$ $>>$ $R_{*}$, as:

\begin{equation}
F^{-}({\nu}_{1},{\nu}_{2})\;{\approx}\;  \frac{2}{3}\,\left(\frac{R_{*}}{D}\right)^{3} \left({\int}_{{\nu}_{1}}^{{\nu}_{2}}\frac{B_{\nu}(T_{*})}{h{\nu}} \left[h{\nu}\,-h{\nu}_{1}\right]\,d{\nu}\right)
\end{equation}

\noindent If there is no gas in the system, then all the incident stellar flux heats 
the dust so that  ${\nu}_{1}$ = 0 and ${\nu}_{2}$ = ${\infty}$. In a system with both 
dust and gas where, for simplicity, it is assumed that all gas atoms have the 
same average ionization potential $h{\nu}_{I}$, then the grains absorb all the photons 
with ${\nu}$ $<$ ${\nu}_{I}$ while the gas absorbs all the photons with 
${\nu}$ $>$ ${\nu}_{I}$. Therefore, the flux heating the dust, $F^{-}_{dust}$, is 
computed from Equation (1) for ${\nu}_{1}$ = 0 and ${\nu}_{2}$ = ${\nu}_{I}$.
Since we consider environments where $h{\nu}_{I}$ $>>$ $kT_{*}$, most stellar 
photons are absorbed by grains, and the expected dust temperature is nearly the 
same as if no gas was present. For the flux heating the gas, $F^{-}_{gas}$, we use 
Equation (1) with ${\nu}_{1}$ = ${\nu}_{I}$ and ${\nu}_{2}$ = ${\infty}$. 
Using the Wien approximation for the Planck curve we find:

 \begin{equation}
F^{-}_{gas}\;{\approx}\;\frac{2}{3}\,\left(\frac{R_{*}}{D}\right)^{3}\,\frac{2\,h}{c^{2}}\,\left(\frac{k\,T_{*}}{h}\right)^{4}{\int}_{x_{I}}^{\infty}(x^{3}\,-x_{I}\,x^{2})\,e^{-x}\,dx
\end{equation}

\noindent where:

\begin{equation}
x_{I}\;=\;\frac{h{\nu}_{I}}{k\,T_{*}}.
\end{equation}

\noindent From expressions (1) and (2), we find:

 \begin{equation}
\frac{F^{-}_{gas}}{F^{-}_{dust}}\;{\approx}\;\frac{\left(x_{I}^{2}\;+\;4\,x_{I}\;+\;6\right)e^{-x_{I}}}{{\pi}^{4}/15}.
\end{equation}

The value of $h\nu_I$ is determined from the metallic constituents in a
gaseous disk. Hence, knowledge of the gas disk composition is imperative in 
calculating this average energy of a photoelectron and the distribution of its energy
into the gaseous material. \citet{gaensicke06,gaensicke07,gaensicke08} report
emission lines from Ca in all three white dwarfs and Fe in SDSS1228 and Ton 345, 
absorption lines of Mg in all three white dwarfs, and Ca and Si absorption in Ton 345. 
Under the assumption that heavy
elements in the photospheres of these white dwarfs were accreted from their
disks \citep[see e.g.,][]{zuckerman07,klein10}, we can claim that each gas disk
holds at least the elements Mg, Fe, Si, and Ca. 
With this mix of elements we calculate an average ionizing potential of $\sim$8 eV. 
Better knowledge of the gas disk
composition, and ultimately the energetics of the gas disk, is most likely
to come from deep, high-spectral resolution 
spectroscopic observations of these white dwarfs in the ultraviolet. 
For the case of a star
with effective temperature $T_{*}$ = 20,000 K and $h\nu_I$ = 8 eV,
we compute  $F^{-}_{gas}/F^{-}_{dust}$ ${\approx}$  5\%.

From the measurements reported in Table \ref{tablineflux} for SDSS1228,
Ton 345, and SDSS1043 we see that the line fluxes
received at Earth summed over the calcium triplet 
emission are roughly 10 ${\times}$ 10$^{-15}$, 4 ${\times}$ 10$^{-15}$, and 
2 ${\times}$ 10$^{-15}$ erg cm$^{-2}$ s$^{-1}$, respectively.  For these 
same three stars, the fluxes from the dust
can be very roughly approximated as ${\nu}F_{\nu}$ evaluated at 
3.6 ${\mu}$m, or 2 ${\times}$ 10$^{-13}$, 2 ${\times}$ 10$^{-13}$, and 
3 ${\times}$ 10$^{-14}$ erg cm$^{-2}$ s$^{-1}$. Therefore, for each
of the three white dwarfs, the ratio of the flux received in the calcium triplet line emission 
compared to the flux received from the dust is $\sim$5\%, 
as predicted. We take this agreement of the data with the 
model as support of our description of the systems as Z II regions.

The flux re-radiated by the gas in the disk, $F^{+}_{gas}$ is:

\begin{equation}
F^{+}_{gas}\;=\;{\sum}\,{\pi}\,B_{\nu}(T_{gas})\,{\Delta}{\nu}
\end{equation}

\noindent In this expression, the sum is performed over all important cooling lines, 
here assumed to be 
the three members of the calcium triplet.  Although the disk is orbiting rapidly, we 
assume that each portion of the gas is in vertical hydrostatic equilibrium and that 
the local microscopic gas motions are mainly thermal.   
Therefore:

\begin{equation}
{\Delta}{\nu}\;=\;{\beta}_{line}\,\left(\frac{2\,k\,T_{gas}}{{m(Ca)}}\right)^{1/2}\,\frac{{\nu}_{L}}{c}
\end{equation}

\noindent where ${\nu}_{L}$ is the line frequency, $m(Ca)$ is the mean atomic 
weight of calcium, and ${\beta}_{line}$ is a coefficient of order unity.  We compute
the gas temperature by finding the value of $T_{gas}$ such that:

\begin{equation}
F^{+}_{gas}(T_{gas})\;=\;F^{-}_{gas}.
\end{equation} 

With considerable simplifications, we compute an analytic solution to Equation (6).  
We assign the same frequency to all three members of the calcium triplet and 
take ${\beta}_{line}$ = 4.  Combining the previous equations we find:

 \begin{equation}
F^{+}_{gas}\;{\approx}\;12\,\left(\frac{2\,k\,T_{gas}}{{m(Ca)}}\right)^{1/2}\,\frac{2\,{\pi}\,h\,{\nu}_{L}^{4}}{c^{3}}\,e^{-h{\nu}_L/kT_{gas}}.
\end{equation}

\noindent For convenience, we define $A$ such that:

\begin{equation}
A\;=\;\frac{1}{18\,{\pi}}\left(\frac{R_{*}}{D}\right)^{3}\,c\,\left(\frac{{m(Ca)}}{2\,k\,T_{gas}}\right)^{1/2}\,\left(\frac{{\nu}_{*}}{{\nu}_{L}}\right)^{4}\,\left(x_{I}^{2}\,+\,4\,x_{I}\,+\,6\right)
\end{equation}

\noindent  where $h{\nu}_{*}$ = $k\,T_{*}$.  Combining the above equations, we find:

\begin{equation}
T_{gas}(D)\;=\;T_{*}\,\left(\frac{{\nu}_{I}}{{\nu}_{L}}\,-\,\frac{{\nu}_{*}}{{\nu}_{L}}\,\ln A\right)^{-1}.
\end{equation}

\begin{figure}
 \begin{center}
 \includegraphics[width=160mm]{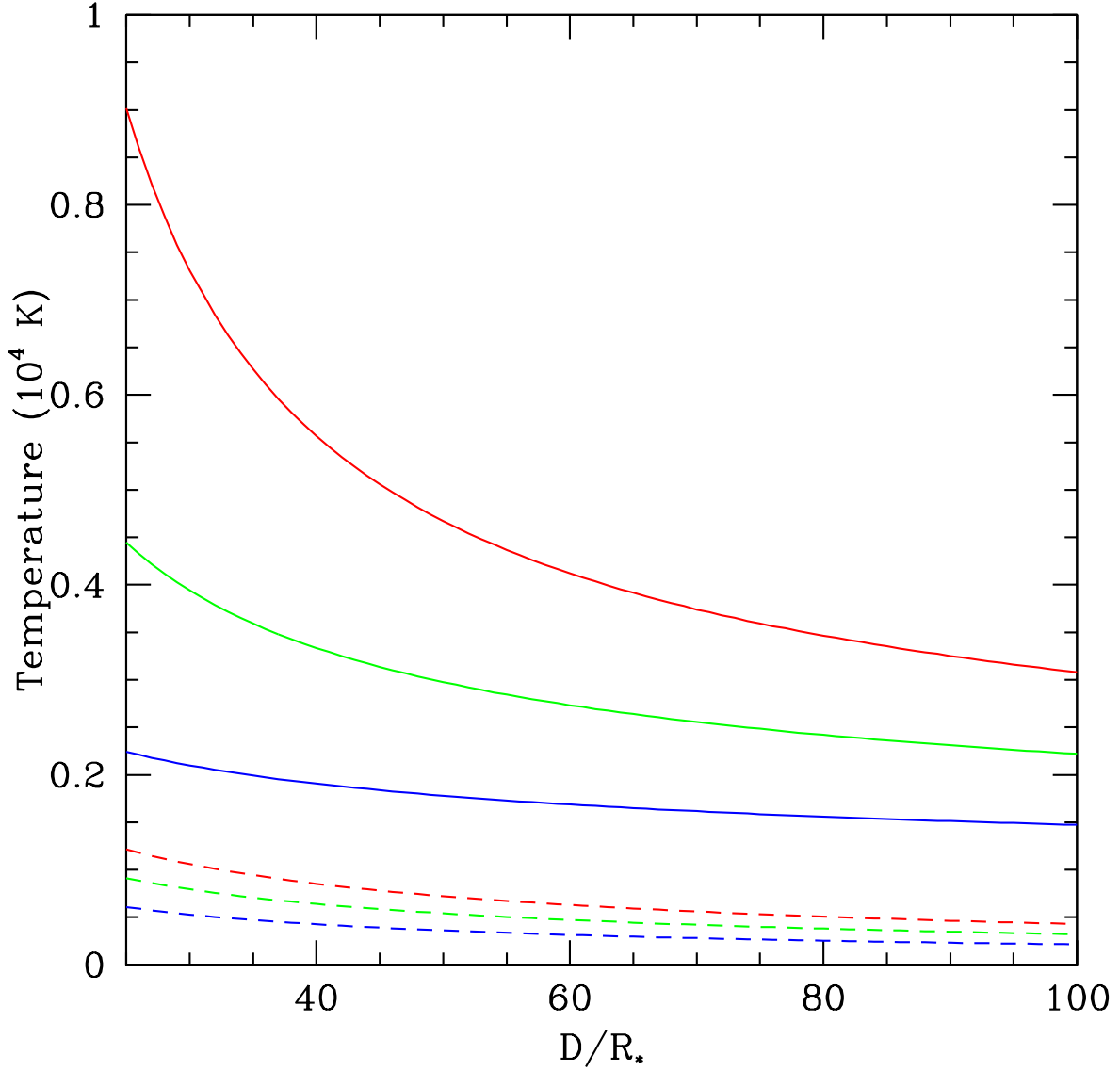}
 \end{center}
\caption{\label{figmodel} Disk temperature as a function of radial separation
from its host white dwarf star in units of stellar radii. The solid lines are for
the gas disk temperature as calculated using
the analytical Z~II region model outlined in Section \ref{seczii}. Dashed
lines are dust disk temperatures as described in Section \ref{secdust}.
The red, green, and blue curves are for  
stellar effective temperatures 
of 20,000 K, 15,000 K, and 10,000 K respectively. We note that the gas disk model
predictions fail for D/R$_*$ $<$ 20 (see Section \ref{seczii}).}
\end{figure}

Gas temperatures are calculated as a function of distance from the star
using the analytic solution in Equation (10) and are plotted for specific cases
in Figure \ref{figmodel}. We assume in these calculations that
$m(Ca)$ = 40 atomic mass units. For $D$ $<<$ 20 $R_{*}$, 
the simple model fails because there is insufficient cooling.  However, for 
$D$ $>>$ $R_{*}$ and T$_{eff}$ = 20,000 K, we see that the gas temperature 
ranges from 6000 K to 3000 K, depending upon the exact distance from the star.  
Such gas disk temperatures are roughly in agreement with those estimated 
from observations of the three gas disk-hosting stars.

\subsubsection{Decoupled Gas and Dust Disks}
\label{sec611}

In our scenario for tidally-disrupted  disks orbiting white dwarfs,  the gas 
temperature and the dust temperature are distinct because the gas and 
dust are spatially separated.   The dust is assumed to be confined to a 
flat disk somewhat analogous to Saturn's ring.  Depending upon the mass 
of the tidally-disrupted rocky object and the extent of the radial zone where 
the material is located, the vertical thickness of the dust may be only 1 cm. 
In contrast, the gas is pictured to be in vertical hydrostatic equilibrium and therefore:

\begin{equation}
{\rho}_{gas}\;=\;{\rho}(0)\,e^{-z^{2}/H^{2}}
\end{equation}

\noindent where ${\rho}(0)$ is the midplane density. For gas orbiting at radial 
distance $D$ from the star of mass $M_{*}$:

\begin{equation}
H\;=\;\left(\frac{2\,k\,T_{gas}\,D^{3}}{G\,M_{*}\,{\mu}}\right)^{1/2}.
\end{equation}

\noindent With $D$ = 10$^{10}$ cm, $T_{gas}$ ${\sim}$ 6,000 K, 
$M_{*}$ = 0.6 M$_{\odot}$, and adopting ${\mu}$ = 2.4 ${\times}$ 10$^{-23}$ g 
as representative of an ionized gas of heavy atoms, we estimate
$H$ ${\approx}$ 3 ${\times}$ 10$^{7}$ cm. Thus, the bulk of the gas is very far 
from the dust and their temperatures may be different. 

\section{Discussion}

From the imaging and spectroscopic data we have estimated inner and outer 
radii for both the gas and dust disks using as few assumptions as possible to 
connect the dust and gas disk results. Now, the two disk parameter sets are combined
to understand how the gas and dust are related, and what that can
tell one about their origin and evolution.

Dust disk inner radii are close to what is expected to correspond
to the sublimation temperature for silicate dust particles.
Spitzer IRS observations 
presented in \citet{jura09} suggest white dwarfs with circumstellar disks
have orbiting dust composed of olivine. The limited wavelength coverage 
of these spectra preclude more complete compositional analyses
\citep[but see][]{reach09}. None of the gas disk-hosting white dwarfs discussed
herein have existing mid-infrared spectroscopic observations, hence
there is some uncertainty in what dust species orbit these three
white dwarf stars and at what temperature these materials will sublimate.
The dust outer disk radius for SDSS1228 is consistent with
the expectation for its Roche radius and in light of similar results
for several other dust disk-hosting white dwarfs \citep[e.g.,][]{farihi09},
we assume the same is true for Ton 345 and SDSS1043.

Gas disk outer radii are spatially coincident (to within measured and modeled
uncertainties) with dust disk outer radii (Figure \ref{fig7gas}). The gas
disk inner radius for SDSS1228 and Ton 345 appears to be farther from the host star
than the dust disk inner radius. For SDSS1043 the gas disk is
closer to the host star, although the uncertainty range for its gas disk inner radius
is large (Table \ref{tabgasdi}). Slightly different
disk inclinations (that are still within the range of viable model parameters,
see Table \ref{tabtonmod}) could push the dust and gas disk inner radii
into better agreement. Lower disk inclination values than those quoted
in Table \ref{tabtonmod} would result in gas disk inner radii closer to
the host white dwarf star and dust disk inner radii farther away from the host
white dwarf star and vice-versa.

Similar outer radii of the dusty and gaseous disks for the three
white dwarfs can be understood through the formalism developed in \citet{jura08}.
\citet{jura08} suggests that pre-existing dusty disk material can sputter
entering rocky object material from a white dwarf's remnant planetary system.
The pre-existing material would have been some rocky material
that was tidally shredded after falling within the white dwarf's tidal radius.
New infalling material would tidally shred and then sputter out onto the
pre-exisiting disk, populating the disk with gaseous material. Steady state
flows of rocky material could maintain the dusty and gaseous
disks that are observed around each of the three white dwarfs.

The above discussion of disk inner radii suggests that the gas and
dust disks are likely to be entirely spatially coincident from inner to outer disk. 
However, the best fit model parameters (Table \ref{tabtonmod}) 
for SDSS1228's and Ton 345's disk systems would have to be stretched to
the extremes of their viable ranges to be in agreement. 
Here we discuss another possible way that the 
dust disks around SDSS1228 and Ton 345 could reside
farther from the host stars and be more in agreement with the gas disk inner radii
without changing disk inclination angles. Flat dust disk 
models predict the temperatures shown in Table \ref{tabtonmod}. 
As suggested by \citet{brinkworth09}, if a warp were present in a 
dust disk then its grains could be cooler and located farther from their 
host white dwarf star while still reproducing the observed excess infrared 
emission. Warps for the dust disk-hosting white dwarfs GD 362 and GD 56
have been suggested by \citet{jura07a,jura09}. 
A slight warp in the dust disk of SDSS1228 
or Ton 345 could inflate the inner radius by a factor
$\gtrsim$1 \citep{jura09}. Such an inflation 
could bring SDSS1228's and Ton 345's inner dust disk radii 
into agreement with their inner gas disk radii. For the case of SDSS1043, 
the combined uncertainties for its gas and dust disk inner radii are 
compatible with both terminating at the same inner radius. The 
asymmetric gas emission lines observed (Figure \ref{fig1gas})
suggest that the white dwarfs' gas disks are not likely to be flat, lending support 
to a warped dust disk interpretation.

Increased disk viscosity at the gas and dust disk inner termination 
radius can provide an explanation for similar gas and dust disk 
inner radii as well as the lack of emission corresponding to material 
extending inwards to the white dwarf stars. In the  
region of the disk where dust grains are not yet sublimated, 
the net viscosity of the disk is
small and disk material does not drift inwards rapidly.
The inner disk is heated to the sublimation temperature where
grains begin to vaporize resulting in inner disk material that is all
gaseous and ionized. Higher densities of hot gas leads to
enhanced disk viscosity which causes 
material to drift inwards towards the host white dwarf
star on short timescales.  
Rapid inner disk depletion will result in small values of
inner disk surface density, consistent with the lack of any
emission having velocities corresponding to radii closer to the white dwarf than the
observed inner disk radii. Thus, the sublimation of dust grains
and resulting enhanced gas density can potentially 
explain coincident gas and dust disk inner radii around these white dwarfs. 
The eventual fate of the inner disk material
will be to accrete onto the host white dwarf star, polluting its
atmosphere and providing a unique insight into its bulk composition
\citep[e.g.,][]{zuckerman07,klein10,dufour10}.

\section{Conclusion}

We have obtained a suite of ground-based optical and near-infrared
measurements of three gas-disk hosting white dwarfs. Unambiguous infrared
excess emission is identified towards Ton 345, confirming that this white dwarf
hosts a dusty disk in addition to its gaseous disk. Characterization of
these three gaseous and dusty disk systems indicates that the gas and dust disks
are spatially coincident at their outer radii and likely their
inner radii. Disk parameterization results are consistent with a scenario
where both dust and gas disks have their origin in the dissolution
of rocky objects from the white dwarfs' remnant planetary systems.

Detection of the Ca~II H \& K and Ca~II IRT lines in emission towards
SDSS1228 enables the development of a new model for the gas disk heating
mechanism. This model relies on photoionization of metallic atoms in a
metal-dominiated region around the hot white dwarfs and cooling of the
gaseous material through optically thick emission lines.



\acknowledgments

C.M. acknowledges support from the Spitzer Visiting Graduate Student program and
from a LLNL Minigrant to UCLA.
We would like to thank the observers and queue coordinators who carried out
service observations at CFHT (programs 08AD96). We thank Detlev Koester for
providing atmospheric models for SDSS1228 and SDSS1043
and Jay Farihi for helpful discussion. We thank Carolyn Brinkworth for useful discussion
and for allowing us to use unpublished Spitzer results for SDSS1043.
Based on observations obtained with WIRCam, a joint project of CFHT, Taiwan, Korea, 
Canada, France, at the Canada-France-Hawaii Telescope (CFHT) which is operated by 
the National Research Council (NRC) of Canada, the Institute National des Sciences 
de l'Univers of the Centre National de la Recherche Scientifique of France, and the
University of Hawaii. 
Based on observations obtained at the Gemini Observatory, which is operated by the
Association of Universities for Research in Astronomy, Inc., under a cooperative agreement
with the NSF on behalf of the Gemini partnership: the National Science Foundation (United
States), the Science and Technology Facilities Council (United Kingdom), the
National Research Council (Canada), CONICYT (Chile), the Australian Research Council
(Australia), MinistŽrio da Cincia e Tecnologia (Brazil) 
and Ministerio de Ciencia, Tecnolog'a e Innovaci—n Productiva  (Argentina).
We are grateful to the Director of the Gemini North telescope for granting us
observing time.
Some of the data presented herein were obtained at the W.M. Keck Observatory, 
which is operated as a scientific partnership among the California Institute of 
Technology, the University of California and the National Aeronautics and Space 
Administration. The Observatory was made possible by the generous financial 
support of the W.M. Keck Foundation.
This publication makes use of data products from the Two Micron All Sky
Survey, which is a joint project of the University of Massachusetts and the
Infrared Processing and Analysis Center/California Institute of Technology,
funded by the National Aeronautics and Space Administration and the National
Science Foundation. Funding for the SDSS and SDSS-II has been provided by the Alfred P. Sloan Foundation, the Participating Institutions, the National Science Foundation, the U.S. Department of Energy, the National Aeronautics and Space Administration, the Japanese Monbukagakusho, the Max Planck Society, and the Higher Education Funding Council for England.
Based on observations made with the NASA 
Galaxy Evolution Explorer. GALEX is operated for NASA by the California 
Institute of Technology under NASA contract NAS5-98034.
This research was supported in part by NASA and NSF grants to UCLA.



{\it Facilities:} \facility{CFHT (WIRCam)}, \facility{Gemini:North (NIRI)}, \facility{Keck:I (HIRES)},
\facility{Shane (Gemini)}




\clearpage

\clearpage

\clearpage

\clearpage

\clearpage

\clearpage

\clearpage

\clearpage

\clearpage

\clearpage

\begin{deluxetable}{cccccccc}
\tabletypesize{\small}
\tablecolumns{8}
\tablewidth{0pt}
\tablecaption{Broad-band Fluxes for Ton 345 \label{tabtonflux}}
\tablehead{ 
  \colhead{Band} & 
  \colhead{$\lambda$} & 
  \multicolumn{3}{c}{mag} &
  \multicolumn{3}{c}{Flux Density} \\
  \colhead{} &
  \colhead{nm} &
  \multicolumn{3}{c}{}&
  \multicolumn{3}{c}{(mJy)}
}
\startdata
\multicolumn{8}{c}{NIR} \\
                        &            &   WIRCam\tablenotemark{b} &  NIRI\tablenotemark{b} &   Average   & F$_{\rm obs}$ & F$_{*}$\tablenotemark{a} & F$_{\rm excess}$ \\
\tableline
L$^{\prime}$ & 3761 &     $-$        &  15.2$\pm$0.3  & $-$            & 0.21$\pm$28\% & 0.075 & 0.135$\pm$0.059  \\
K$_{\rm s}$ & 2157 & 15.95$\pm$0.03 & 16.04$\pm$0.03 & 15.99$\pm$0.02 & 0.27$\pm$2\%  & 0.19  &  0.076$\pm$0.005 \\
H & 1654  & 16.18$\pm$0.05 & 16.33$\pm$0.02 & 16.26$\pm$0.07 & 0.33$\pm$7\%   &  0.29 &  0.036$\pm$0.023  \\
J & 1215   & 16.36$\pm$0.03 & 16.41$\pm$0.03 & 16.39$\pm$0.02 & 0.45$\pm$2\%   & 0.45  & 0.00   \\
\cutinhead{SDSS DR7$-$ Optical}
{\it z} &  893.1    & \multicolumn{3}{c}{16.615$\pm$0.018} & \multicolumn{3}{c}{0.82$\pm$1.8\%} \\
{\it i} &  748.1    & \multicolumn{3}{c}{16.348$\pm$0.015} & \multicolumn{3}{c}{1.05$\pm$1.5\%} \\
{\it r} &  616.5    & \multicolumn{3}{c}{16.077$\pm$0.014} & \multicolumn{3}{c}{1.35$\pm$1.4\%} \\
{\it g} &  468.6    & \multicolumn{3}{c}{15.728$\pm$0.035} & \multicolumn{3}{c}{1.86$\pm$3.5\%} \\
{\it u} &  355.1    & \multicolumn{3}{c}{15.559$\pm$0.015} & \multicolumn{3}{c}{2.25$\pm$1.5\%} \\
\cutinhead{GALEX\tablenotemark{b} $-$ Ultraviolet} 
NUV  &     227.1   &  \multicolumn{3}{c}{15.50$\pm$0.03}   &   \multicolumn{3}{c}{2.28$\pm$3\%} \\
FUV  &     152.8   &  \multicolumn{3}{c}{15.69$\pm$0.05}   &   \multicolumn{3}{c}{1.92$\pm$5\%} \\
\enddata
\tablenotetext{a}{F$_{*}$ is the predicted photospheric flux density in the given bandpass assuming Ton 345 is a blackbody with T$_{\rm eff}$ of 18,600 K and that the J-band monochromatic flux is entirely photospheric in nature.}
\tablenotetext{b}{WIRCam and NIRI magnitudes are in the Johnson system. GALEX measurements are in AB magnitudes. GALEX uncertainties are as suggested in
\citet{morrissey07}.}
\end{deluxetable}

\clearpage

\begin{deluxetable}{cccc}
\tabletypesize{\normalsize}
\tablecolumns{4}
\tablewidth{0pt}
\tablecaption{Broad-band Fluxes for SDSS1043 \label{tab10flux}}
\tablehead{ \colhead{Band} & \colhead{$\lambda$} & \colhead{mag} & \colhead{F$_{\rm obs}$} \\
                      \colhead{}           & \colhead{nm}              & \colhead{}         & \colhead{(mJy)}
}
\startdata
\cutinhead{WIRCam/Gemini\tablenotemark{a,b} \ $-$ NIR} \\
K$_{\rm s}$ & 2157  & 17.69$\pm$0.05 & 0.056$\pm$5\%  \\
J                    & 1210  & 17.72$\pm$0.14 & 0.133$\pm$13\% \\
\cutinhead{SDSS DR7 $-$ Optical} 
{\it z} &  893.1   & 17.942$\pm$0.036 & 0.24$\pm$3.6\% \\
{\it i} &  748.1    & 17.701$\pm$0.017 & 0.30$\pm$1.7\% \\
{\it r} &  616.5    & 17.433$\pm$0.014 & 0.39$\pm$1.4\% \\
{\it g} &  468.6   & 17.123$\pm$0.022 & 0.51$\pm$2.2\% \\
{\it u} &  355.1   & 17.358$\pm$0.022 & 0.43$\pm$2.2\% \\
\cutinhead{GALEX\tablenotemark{b} $-$ Ultraviolet} 
NUV  &     227.1   &  17.07$\pm$0.03 &   0.54$\pm$3\% \\
FUV  &     152.8   &  16.86$\pm$0.05  &   0.66$\pm$5\% \\
\enddata
\tablenotetext{a}{K$_{\rm s}$-band data are from CFHT WIRCam observations while
J-band data are from Lick Gemini observations (see Section
\ref{secobs}).}
\tablenotetext{b}{WIRCam and Gemini magnitudes are in the Johnson system. GALEX measurements are in AB magnitudes. GALEX uncertainties are as suggested in
\citet{morrissey07}.}
\end{deluxetable}

\clearpage

\begin{deluxetable}{ccc}
\tabletypesize{\normalsize}
\tablecolumns{3}
\tablewidth{0pt}
\tablecaption{NIRI Fluxes For Background Galaxy Companion to Ton 345 \label{tabgalflux}}
\tablehead{ \colhead{Band} & \colhead{mag} & \colhead{Flux Density} \\
\colhead{} & \colhead{} & \colhead{($\mu$Jy)} }
\startdata
J           & $>$21.0 & $<$6.5 \\
H           &    19.15$\pm$0.24 & 22.9$\pm$22\% \\
K$_{\rm s}$ &    18.47$\pm$0.15 & 27.3$\pm$14\% \\
\enddata
\tablecomments{The background galaxy at 1.8$^{\prime\prime}$ separation from Ton 345 appears to
be a J-band drop-out galaxy. Comparing to galaxy models 
\citep[e.g., see Figure 11 of][and references therein]{sanders07} suggests that this is a Balmer-break
Spiral or Cool ULIRG galaxy at a redshift $z$$\sim$2.}
\end{deluxetable}

\clearpage

\begin{deluxetable}{lcccc}
\rotate
\tabletypesize{\normalsize}
\tablecolumns{6}
\tablewidth{0pt}
\tablecaption{HIRES Observations Summary \label{tabhobs}}
\tablehead{
\colhead{UT Date} & \colhead{Setup} & \colhead{Coverage} & \colhead{Integration Time (sec)} & \colhead{S/N\tablenotemark{a}} 
}
\startdata
\multicolumn{5}{l}{\bf SDSS1228} \\
05 May 2007 & UV Collimator  & 3120-5950 \AA\ & 3000 & 22 \\
14 Nov 2008 & Red Collimator & 4500-9000 \AA\ & 1800 & 12 \\
15 Nov 2008 & Red Collimator & 4500-9000 \AA\ & 1600 & 13 \\
16 Nov 2008 & Red Collimator & 4500-9000 \AA\ & 1900 & 12 \\
\multicolumn{5}{l}{\bf Ton 345} \\
13 Feb 2008 & UV Collimator  & 3130-5960 \AA\ &  2$\times$1800 & 38\tablenotemark{b} \\
14 Feb 2008 & UV Collimator  & 3130-5960 \AA\ &  2$\times$1500 & 38\tablenotemark{b} \\
26 Feb 2008 & Red Collimator & 4600-9150 \AA\ & 3600 & 42 \\
14 Nov 2008 & Red Collimator & 4500-9000 \AA\ & 2$\times$1800 & 45\tablenotemark{b} \\
16 Nov 2008 & Red Collimator & 4500-9000 \AA\ & 1800 & 32 \\
\multicolumn{5}{l}{\bf SDSS1043} \\
05 May 2007 & UV Collimator  & 3120-5950 \AA\ & 3000 & 13 \\
15 Nov 2008 & Red Collimator & 4500-9000 \AA\ & 2$\times$2000 & 18\tablenotemark{b} \\
16 Nov 2008 & Red Collimator & 4500-9000 \AA\ & 1500 \& 1700 & 14\tablenotemark{b} \\
\enddata
\tablecomments{A resolving power of $\sim$40,000 was measured for all epochs from the FWHM of single arclines in comparison spectra.}
\tablenotetext{a}{S/N measured at 5750 \AA .}
\tablenotetext{b}{S/N for combined exposures.}
\end{deluxetable}

\clearpage

\begin{deluxetable}{lcccccc}
\rotate
\tabletypesize{\scriptsize}
\tablecolumns{7}
\tablewidth{0pt}
\tablecaption{Gas-disk Emission Line Measurements \label{tabgas}}
\tablehead{
 \colhead{Transition} & 
 \colhead{UT Date\tablenotemark{a}} & 
 \colhead{Equivalent Width} & 
 \colhead{Peak Separation\tablenotemark{b}} & 
 \colhead{Full Width\tablenotemark{c}} & 
 \colhead{Peak Midpoint Velocity\tablenotemark{b}} & 
 \colhead{v$_{\rm max}$sin$i$\tablenotemark{b}} \\
 \colhead{} & 
 \colhead{} & 
 \colhead{(\AA )}           & 
 \colhead{(km s$^{-1}$)} & 
 \colhead{(km s$^{-1}$)} & 
 \colhead{(km s$^{-1}$)} & 
 \colhead{(km s$^{-1}$)} 
}
\startdata
\multicolumn{7}{l}{\bf SDSS 1228} \\
Ca~II K                              & 05 May 2007    & 0.6$\pm$0.1 & 707$\pm$20 & 1014$\pm$44 & -35$\pm$10 & 455$\pm$23/559$\pm$38 \\
Ca~II H                              & 05 May 2007    & $-$\tablenotemark{f} & 731$\pm$18 & 1020$\pm$53 & -22$\pm$9 & 489$\pm$38/531$\pm$38\\
\\*[-0.05in]
\multirow{2}{*}{Fe~II $\lambda$5169}\tablenotemark{g} & 05 May 2007    & 1.3$\pm$0.1 & $-$        & $-$ & $-$ & $-$ \\
                                     & 14,15,16 Nov 2008 & 1.3$\pm$0.1 & $-$     & $-$ & $-$ & $-$ \\
\\*[-0.05in]
Ca~II $\lambda$8498                  & 14,15,16 Nov 2008 & 13.8$\pm$1.0 & 656$\pm$9 & 938$\pm$16 & -22$\pm$4  & 379$\pm$7/560$\pm$14 \\
Ca~II $\lambda$8542                  & 14,15,16 Nov 2008 & 19.3$\pm$1.4 & 655$\pm$9 & 944$\pm$9 & -17$\pm$4 &  393$\pm$7/551$\pm$14 \\
Ca~II $\lambda$8662                  & 14,15,16 Nov 2008 & 19.8$\pm$1.7\tablenotemark{e} & 659$\pm$5 & 933$\pm$35 & -18$\pm$3 &  382$\pm$7/548$\pm$35 \\
\\*[-0.05in]
\multicolumn{7}{l}{\bf Ton 345} \\
\multirow{2}{*}{Fe~II $\lambda$5169} & 13,14.26 Feb 2008\tablenotemark{d} & 1.1$\pm$0.2 & 838$\pm$14 & 1349$\pm$87 & -79$\pm$14  & $-$ \\ 
                                                                    & 14,16 Nov 2008                                      & 1.1$\pm$0.2 & 869$\pm$18 & 1305$\pm$96  & -69$\pm$18  & $-$ \\
\\*[-0.05in]
\multirow{2}{*}{Ca~II $\lambda$8498} & 26 Feb 2008    & 5.2$\pm$0.4 & 842$\pm$20 & 1305$\pm$49 & -90$\pm$11 &  671$\pm$35/634$\pm$35 \\
                                     & 14,16 Nov 2008 & 4.9$\pm$0.4 & 827$\pm$18 & 1252$\pm$73 & -104$\pm$8  & 619$\pm$18/634$\pm$71 \\
\\*[-0.05in]
\multirow{2}{*}{Ca~II $\lambda$8542} & 26 Feb 2008    & 4.1$\pm$0.3 & 857$\pm$16 & 1281$\pm$63 & -98$\pm$8 & 706$\pm$35/575$\pm$53\\
                                     & 14,16 Nov 2008 & 3.5$\pm$0.3 & 891$\pm$13  & 1211$\pm$56 & -89$\pm$6 & 618$\pm$18/593$\pm$53 \\
\\*[-0.05in]
\multirow{2}{*}{Ca~II $\lambda$8662} & 26 Feb 2008    & 3.4$\pm$0.2 & 876$\pm$17 & 1298$\pm$75 & -99$\pm$9 & 663$\pm$28/635$\pm$69 \\
                                     & 14,16 Nov 2008 & 3.4$\pm$0.5\tablenotemark{e} & 882$\pm$25 & 1263$\pm$71 & -89$\pm$12 &  611$\pm$17/652$\pm$69\\
\\*[-0.05in]
\multicolumn{7}{l}{\bf SDSS1043} \\
Ca~II $\lambda$8498                  & 15,16 Nov 2008 & 7.0$\pm$0.5 & 757$\pm$38 & 1482$\pm$50 & 39$\pm$19  & 786$\pm$35/686$\pm$35 \\
Ca~II $\lambda$8542                  & 15,16 Nov 2008 & 7.0$\pm$0.6 & 778$\pm$49 & 1439$\pm$50 & 52$\pm$25 & 811$\pm$35/628$\pm$35 \\
Ca~II $\lambda$8662                  & 15,16 Nov 2008 & 8.1$\pm$1.2\tablenotemark{e} & 806$\pm$48 & 1402$\pm$109 & 42$\pm$24 & 802$\pm$104/600$\pm$35 \\
\enddata
\tablenotetext{a}{Spectra have been combined when multiple days are listed.}
\tablenotetext{b}{See Sections \ref{secspec} and \ref{secgasdiskdi} 
for description of these quantities. The two different values reported for v$_{\rm max}$sin$i$
correspond to the maximum velocity gas seen in the blue and red wings of the double-peaked
emission features, respectively.}
\tablenotetext{c}{Full velocity width of emission feature, from continuum blueward of the blue emission peak to continuum redward of the red emission peak.}
\tablenotetext{d}{To within the noise of each individual exposure the 13,14 Feb 2008 and 26 Feb 2008 Fe~II $\lambda$5169 emission features for Ton 345 are identical; to improve the S/N we combine all three spectra into a single 2008 Feb epoch.}
\tablenotetext{e}{Measurement may be contaminated by second order blue light leak, see Section \ref{sechobs}.}
\tablenotetext{f}{Photospheric Balmer H$\epsilon$ absorption in SDSS1228 prevents an accurate continuum estimation for the Ca~II H emission complex.}
\tablenotetext{g}{The Fe~II $\lambda$5169 emission in SDSS1228 shows peculiar features (see Figure \ref{fig4gas}), the origin of which are uncertain. As such we do not include these measurements in this table.}
\end{deluxetable}

\clearpage

\begin{deluxetable}{lcccccccccc}
\rotate
\centering
\tabletypesize{\footnotesize}
\tablecolumns{11}
\tablewidth{0pt}
\tablecaption{Kinematics of Gas-disk White Dwarfs \label{tabkin}}
\tablehead{
 \colhead{WD Name} & 
 \colhead{RV$_{\rm obs}$\tablenotemark{a}} & 
 \colhead{M$_{\rm WD}$} & 
 \colhead{R$_{\rm WD}$} & 
 \colhead{v$_{\rm grav}$\tablenotemark{b}} & 
 \colhead{Distance} & 
 \colhead{pmRA} &
 \colhead{pmDE} & 
 \colhead{U} & 
 \colhead{V} & 
 \colhead{W} \\
 \colhead{} & 
 \colhead{(km s$^{-1}$)}  & 
 \colhead{(M$_{\odot}$)}& 
 \colhead{(R$_{\odot}$)}& 
 \colhead{(km s$^{-1}$)}  & 
 \colhead{(pc)}     & 
 \colhead{(mas yr$^{-1}$)} & 
 \colhead{(mas yr$^{-1}$)} & 
 \colhead{(km s$^{-1}$)} & 
 \colhead{(km s$^{-1}$)} & 
 \colhead{(km s$^{-1}$)}
}
\startdata
SDSS1228 & +37$\pm$4 & 0.77 & 0.011 & +44 & 125$\pm$13 & $-$44.0$\pm$2.0 & $-$18.0$\pm$1.0 & $-$18 & $-$20 & $-$12 \\
Ton 345 & +36$\pm$4 & 0.70 & 0.010 & +44 & 85$\pm$10 & $-$10.0$\pm$1.0 & $-$62.0$\pm$2.0 & +9 & $-$21 & $-$14 \\
SDSS1043 & +39$\pm$4 & 0.67 & 0.012 & +35 & 185$\pm$19 & +38.0$\pm$6.0 & $-$46.0$\pm$4.0 & +45 & $-$27 & +6 \\
\enddata
\tablenotetext{a}{RV$_{\rm obs}$ is the observed radial velocity as measured from the absorption lines of each white dwarf. It has not been corrected for each white dwarf's gravitational redshift.}
\tablenotetext{b}{Gravitational redshift as computed from the model-determined mass and radius
reported for each white dwarf in columns 3 and 4.}
\tablecomments{See Section \ref{secspec} for discussion of input and output parameters.
For the UVW space motions positive U is towards the Galactic center, positive V 
is in the direction of Galactic rotation, and positive W is toward the north Galactic pole.}
\end{deluxetable}

\clearpage

\begin{deluxetable}{cccccc}
\tabletypesize{\normalsize}
\tablecolumns{6}
\tablewidth{0pt}
\tablecaption{Dust Disk Models \label{tabtonmod}}
\tablehead{
    \colhead{} &
    \multicolumn{3}{c}{SDSS1228} &
    \colhead{Ton 345} & 
    \colhead{SDSS1043}
    }
\startdata
\cutinhead{Stellar Parameters}
T$_{\rm eff}$ (K) &  \multicolumn{3}{c}{22,020} & 18,600 & 18,330  \\
R$_{\rm WD}$/Dist & \multicolumn{3}{c}{2.0 $\times$ 10$^{-12}$} & 2.6 $\times$ 10$^{-12}$ & 1.5 $\times$ 10$^{-12}$  \\
\cutinhead{Disk Parameters\tablenotemark{a}}
T$_{\rm inner}$ (K) & 1100 & 1300$\pm$50 & 1350 & 1500 & 1200 \\
T$_{\rm outer}$ (K) & 700 & 500$\pm$70 & 350 & 1000 & 1100  \\
$i$ ($^{\circ}$)         & 60 & 73$\pm$3 & 76 &    66  & 60 \\
\enddata
\tablenotetext{a}{The three different parameter sets for SDSS1228 correspond to the
                               three curves plotted in Figure \ref{fig12sed}. The parameters are for the 
                               red, black, and blue
                               curves respectively. We adopt the middle parameter set 
                               (having T$_{\rm inner}$ of 1300 K) for SDSS1228's
                               dust disk properties (corresponding to the black curve in Figure \ref{fig12sed}).
                               From the three curves we estimate the 1$\sigma$ uncertainties on
                               the model parameters (see Section \ref{secdust}) 
                               which are reported with SDSS1228's best fit
                               parameters.}
\tablecomments{The effective temperatures for the three white dwarfs come from 
 \citet{gaensicke06,gaensicke07,gaensicke08}. The ratio of the radius of the white dwarf to the 
 distance of the white dwarf from Earth, R$_{\rm WD}$/Dist; temperature at the
 inner edge of the disk, T$_{\rm inner}$; and inclination angle, $i$, come
 from the model fit to the UV, optical, and infrared photometry.
 See Section \ref{secdust} for a 
 discussion of the temperature at the outer edge of the disk (T$_{\rm outer}$).}
\end{deluxetable}

\clearpage

\begin{deluxetable}{lcccccc}
\tabletypesize{\normalsize}
\tablecolumns{7}
\tablewidth{0pt}
\tablecaption{Disk Dimensions \label{tabgasdi}}
\tablehead{
 \colhead{}            & 
 \multicolumn{3}{c}{Gas Disks} & 
 \multicolumn{3}{c}{Dust Disks} \\
 \colhead{Name} & 
 \colhead{v$_{\rm max}$} & 
 \colhead{R$_{inner,gas}$}  &
 \colhead{R$_{outer,gas}$} &
 \colhead{R$_{inner,dust}$} &
 \colhead{R$_{outer,dust}$} &
 \colhead{T$_{outer}$} \\
 \colhead{}            &   
 \colhead{(km s$^{-1}$)}           &   
 \colhead{(R$_{\rm WD}$)} &
 \colhead{(R$_{\rm WD}$)} &
 \colhead{(R$_{\rm WD}$)} &
 \colhead{(R$_{\rm WD}$)} &
 \colhead{(K)}
}
\startdata
SDSS1228 & 575$\pm$17 & 40$\pm$3 & 108 & 26$\pm$2 & 93$\pm$17 & 500 \\
Ton 345      & 709$\pm$20 & 27$\pm$2 & $\sim$100 & 17$\pm$2 & 100$\pm$18 & 400 \\
SDSS1043 & 923$\pm$52 & 12$\pm$2 & 81 & 23$\pm$2 & 80$\pm$15 & 470 \\
\enddata
\tablecomments{\small{All quoted uncertainties are 1$\sigma$ (see below).
{\it Gas disks:} v$_{\rm max}$ comes from the average of the 
blue or red wing values (whichever is higher) listed 
in the ``v$_{\rm max}$sin$i$'' column of Table \ref{tabgas}. This average
value is then divided by the sine of the inclination angle reported in Table \ref{tabtonmod}.
R$_{inner,gas}$ is derived directly from the value reported in the ``v$_{\rm max}$'' column. 
The quoted uncertainties for R$_{inner,gas}$ also include the statistical uncertainties for
the white dwarf masses and radii as quoted in \citet{gaensicke06,gaensicke07,gaensicke08}.
There may be additional systematic uncertainties that are not included. R$_{outer,gas}$ is
taken from \citet{gaensicke06,gaensicke07,gaensicke08,gaensicke08ASPC}.
{\it Dust disks:} See Section \ref{secdust} for a description of how R$_{inner,dust}$ and
R$_{outer,dust}$ are derived. T$_{outer}$ corresponds to the outer dust radius shown.
The uncertainties for R$_{inner,dust}$ and R$_{outer,dust}$ are estimated from
the range of possible disk parameters modeled for the SDSS1228 data set and the 
suggested uncertainty for each star's effective temperature as quoted in 
\citet{gaensicke06,gaensicke07,gaensicke08}.}}
\end{deluxetable}

\clearpage

\begin{deluxetable}{lccccccc}
\rotate
\centering
\tabletypesize{\footnotesize}
\tablecolumns{7}
\tablewidth{0pt}
\tablecaption{Ca~II Emission Line Flux \label{tablineflux}}
\tablehead{
 \colhead{Name} &
 \colhead{Ca~II K} &
 \colhead{Ca~II $\lambda$8498} &
 \colhead{Ca~II $\lambda$8542} &
 \colhead{Ca~II $\lambda$8662} &
 \colhead{Ratio} &
 \colhead{Gas Disk} \\
 \colhead{} &
 \multicolumn{4}{c}{Total Line Flux\tablenotemark{a} (10$^{-15}$ ergs cm$^{-2}$ s$^{-1}$)} &
 \colhead{F$_{\rm IRT}$/F$_{\rm HK}$} &
 \colhead{Temperature (K)}
}
\startdata
SDSS1228 & 1.4 & 2.6 & 3.6 & 3.7  & $\sim$3.5 & $\sim$5000 \\
Ton 345      & $<$2.5\tablenotemark{b} & 1.8 & 1.3 & 1.2 & $>$0.9\tablenotemark{b} & $<$6500 \\
SDSS1043 & $<$0.6\tablenotemark{b} & 0.7 & 0.7 & 0.8 & $>$2.0\tablenotemark{b} & $<$6000 \\
\enddata
\tablenotetext{a}{These values are computed by multiplying emission line 
EW measurements reported in Table \ref{tabgas} by the stellar continuum flux 
at the emission line location.}
\tablenotetext{b}{These upper limits are calculated assuming we could not detect Ca~II H \& K
emission features weaker than those seen in SDSS1228 (see Figure \ref{fig1gas}).}
\tablecomments{See Section \ref{secmodels} for discussion.}
\end{deluxetable}



\end{document}